\documentclass[preprint,12pt,authoryear]{elsarticle}

\usepackage{amssymb}
\usepackage{amsmath}
\usepackage{hyperref}
\usepackage{xcolor}
\hypersetup{
    colorlinks=true,   
    linkcolor=blue,    
    citecolor=blue,    
    urlcolor=blue      
}
\usepackage{booktabs}
\usepackage{rotating}
\usepackage{amsfonts}
\usepackage[utf8]{inputenc} 
\usepackage[T1]{fontenc}    
\usepackage{multirow}
\usepackage{float}
\usepackage{rotating}
\usepackage{svg}
\usepackage{longtable}
\usepackage{cleveref}

\journal{Remote Sensing of Environment}

\begin{document}

\begin{frontmatter}

\title{How Usable Are Geospatial Foundation Models? \\A Systematic Evaluation of 89 Models}

\author[cam]{Robin Young}
\ead{ray25@cam.ac.uk}

\author[waterloo]{Artyom Gabtraupov\corref{contrib}}

\author[waterloo]{Kenzy Soror\corref{contrib}}

\author[cam]{Srinivasan Keshav}
\ead{sk818@cam.ac.uk}

\cortext[contrib]{These authors contributed equally.}

\affiliation[cam]{organization={Department of Computer Science and Technology},
            institution={University of Cambridge},
            city={Cambridge},
            postcode={CB3 0FD}, 
            state={Cambridgeshire},
            country={UK}}

\affiliation[waterloo]{organization={David R. Cheriton School of Computer Science},
            institution={University of Waterloo},
            city={Waterloo},
            postcode={N2L 3G1}, 
            state={Ontario},
            country={Canada}}

\begin{abstract}
Geospatial foundation models (GeoFMs) offer transformative potential for environmental monitoring, yet adoption among ecologists is uneven. Most evaluations are model-centric, focusing on architecture and benchmark accuracy, which overlooks whether the systems are usable by their intended audiences. To address this gap, we first conducted a pilot expert elicitation survey with ecology and conservation scientists that helped us identify misalignments between current GeoFM development priorities and their needs. Informed by these findings and based on HCI theory, we created a seven-dimension evaluation covering Access \& Deployment, Interaction \& Customization, Trust \& Transparency, Community \& Support, Scientific Permanence, Multilingual Support, and Offline Usability. Then, two raters applied this rubric to 89 GeoFMs. We found distinct accessibility gaps where nearly a third provide no support to practitioners beyond their source code. Dimensions along which ratings were highly consistent function as field-level diagnostics, revealing where there is room for improvement for current GeoFMs.
\end{abstract}




\end{frontmatter}



\section{Introduction}
\label{sec:introduction}
 
Accelerating climate change, unprecedented biodiversity loss, and widespread environmental degradation demand rapid, data-driven responses that rely on comprehensive Earth observation capabilities \citep{wulder2019current, xiao2025foundationmodelsremotesensing, zhu2024foundationsearthclimatefoundation}. 
The emergence of geospatial foundation models (GeoFMs) has transformed Earth observation by enabling rich, learned representations of satellite imagery that generalize across diverse downstream tasks with minimal task-specific training. 
Recent advances including SatMAE \citep{Cong2022}, ScaleMAE \citep{Reed2023}, Prithvi \citep{jakubik2023foundationmodelsgeneralistgeospatial}, Google AlphaEarth \citep{brown2025alphaearthfoundationsembeddingfield} and Tessera \citep{feng2025tessera, feng2026tesserav2} demonstrate applications ranging from crop stress detection to urban heat mapping. 
Multi-modal approaches \citep{xiong2024neuralplasticityinspiredmultimodalfoundation, Guo2024} and vision-language models \citep{klemmer2025satclip, kuckreja2023geochat, zhang2024earthgpt} extend these capabilities to natural language interaction, while weather and climate foundation models \citep{schmude2024prithviwxcfoundationmodel, bodnar2025aurora, allen2025aardvark} achieve performance comparable to traditional numerical weather prediction at a fraction of the computational cost.
 
However, bridging the gap between this technical sophistication and practical use by domain experts remains a challenge. The scientists who best understand environmental processes, including ecologists monitoring deforestation, city planners assessing climate resilience, and emergency managers coordinating disaster response, are frequently the ones who face the steepest technical barriers to using these tools. Current interfaces assume machine learning expertise, requiring users to navigate complex software environments, understand what high-dimensional embeddings represent and how to adapt them, and implement transfer learning pipelines. Model-centric evaluation reinforces this disconnect by prioritizing benchmark performance, which fails to capture the user-centered dimensions in both technical (data access, environment setup) and sociotechnical (documentation, community support, long-term reproducibility) aspects that determine practical adoption.
 
This gap between technical capability and practical usability reflects a long-standing accessibility challenge in Earth observation \citep{pilone2014usable, volentine2021accessibility}. The 2030 UN Sustainable Development Goals \citep{UNSDG} require monitoring at unprecedented scales \citep{kavvada2020towards}, making geospatial AI accessibility not merely a usability concern but a prerequisite for planetary stewardship.

To address this gap, we present the first dedicated usability-centric evaluation framework for geospatial foundation models. Rather than asking ``How well does the model perform?'', we ask: ``How well does the system empower its intended users?'' Our contributions are threefold:
 
\begin{enumerate}
    \item A \textbf{pilot expert elicitation survey} ($N=11$) with ecology and conservation scientists that grounds and motivates our framework. Rather than deriving evaluation dimensions from theory alone, we use practitioner input to surface where GeoFM development priorities may diverge from domain expert needs.
    \item A \textbf{seven-dimension evaluation framework} grounded in HCI theory, covering Access \& Deployment, Interaction \& Customization, Trust \& Transparency, Community \& Support, Scientific Permanence \& Reproducibility, Multilingual Support, and Offline Usability.
    \item A \textbf{systematic evaluation of 89 geospatial foundation models} by two calibrated raters with measured inter-rater reliability in Cohen's Kappa and Gwet's AC1 metrics, producing a comprehensive landscape assessment of GeoFM usability.
\end{enumerate}

Table~\ref{tab:dimensions_overview} summarizes the framework dimensions; each is developed fully in Section~\ref{sec:framework}.

\begin{table*}[htbp]
\centering
\caption{Overview of the seven evaluation dimensions. Each dimension is developed in Section~\ref{sec:framework}.}
\label{tab:dimensions_overview}
\begin{tabular}{llc}
\toprule
\textbf{Dimension} & \textbf{Guiding Question} & \\
\midrule
D1: Access \& Deployment & How much expertise is needed to complete a basic task? \\
D2: Interaction \& Customization & What interaction paradigms are supported? \\
D3: Trust \& Transparency & What helps users validate and trust outputs?  \\
D4: Community \& Support & What resources exist when users get stuck? \\
D5: Scientific Permanence & Will work built on this model remain reproducible? \\
D6: Multilingual Support & Can users access documentation in multiple languages? \\
D7: Offline Usability & Can users work under low-connectivity conditions? \\
\bottomrule
\end{tabular}
\end{table*}

 
\subsection{Geospatial Foundation Model Surveys}
 
Several recent surveys have cataloged the growing landscape of GeoFMs. \citet{xiao2025foundationmodelsremotesensing} provide a detailed review of foundation models for remote sensing, organizing models by architecture and application domain. \citet{zhu2024foundationsearthclimatefoundation} survey foundation models for Earth and climate science, emphasizing training data and benchmark performance. \citet{lu2025vision} surveyed vision foundation models released between 2021 and 2024, categorizing them by perception level and pretraining methodology. \citet{feng2025tessera} include an appendix cataloging over 80 GeoFMs as context for their pre-computed embedding approach. These surveys share a model-centric analytical lens, evaluating models primarily on architecture, training data, parameter count, and benchmark accuracy.
 
What these surveys do not assess is whether any of these models are practically usable by their intended audiences. Existing surveys do not typically examine the deployment interfaces, documentation quality, trust-building features (such as uncertainty quantification, explainable outputs, and validation against known benchmarks), or community support ecosystems that determine whether a domain expert can easily use a given model. Our work fills this gap by applying an explicitly user-centered evaluation lens to the same model corpus.

\subsection{Accessibility Challenges in Earth Observation}
  
\citet{Ziegler2023} documented usability challenges experienced by domain experts using established tools like QGIS and geopandas, finding that users spent disproportionate time on tasks like aligning datasets across differing spatial and temporal resolutions, identifying correct geospatial operators from large and inconsistently named toolsets, and manually tracking analysis provenance across multiple tools, rather than on scientific analysis itself. Climate datasets like CMIP require navigating eight distinct metadata filter categories before basic data search is possible \citep{wcrp_cmip_data_access}. Web Coverage Service implementations like SoilGrids \citep{isric_wcs} require users to construct HTTP requests for basic data access. At the platform level, \citet{dileo2024selfassessment} proposed a self-assessment framework for EO cloud platforms covering dimensions including documentation, community engagement, sustainability, and multilingual support. Our framework extends this user-centered evaluation approach from platforms to foundation models.
 
These patterns extend beyond geospatial tools. The broader scientific computing literature documents similar accessibility challenges across domains. Programming language choice creates community divisions in ecology, with researchers selecting tools based on initial training rather than task suitability \citep{gao2025rapplication}. \citet{hannay2009} surveyed computational practices across scientific disciplines and found that scientists spend approximately 30\% of their time developing software, yet 90\% are primarily self-taught. This pattern produces fragile workflows highly sensitive to software environment changes. \citet{Wilson2014} documented how the absence of software engineering best practices in scientific computing leads to reproducibility failures, wasted effort, and barriers to collaboration. These findings suggest that the accessibility problems we document in GeoFMs are symptomatic of a broader pattern in which scientific tool developers optimize for technical capability while underinvesting in the user-facing infrastructure that determines practical adoption.
 
While it is tempting to view GeoFMs as just the next iteration of geospatial AI tools within existing platforms, there is an interpretive gap that has no direct precedent in the geospatial usability literature. Traditional tools typically apply a task-specific model to produce a specific output, such as a land cover classification, a change detection map, or a spectral index. In contrast, a GeoFM is a general-purpose model pretrained on planetary-scale data. It produces high-dimensional embeddings that are rich, semantic representations of any location on Earth that can be rapidly generalized to diverse downstream tasks. This transforms the core challenge for a domain expert from running a tool to interacting with an abstract data representation, which is a new problem at the intersection of HCI and geospatial machine learning.
 
\subsection{Usability Evaluation in Scientific Software}
 
Our work draws on a broader tradition of usability evaluation using standard HCI techniques applied to scientific and technical software.
 
\citet{swedlow2009} examined usability evaluation methods for scientific software and found that standard HCI evaluation techniques (such as heuristic evaluation, cognitive walkthroughs, and user testing) require adaptation for expert domains where the ``typical user'' (such as scientists) possesses deep specialist knowledge that evaluators (HCI experts) may lack. Their work highlights the importance of domain-grounded evaluation criteria, a principle that motivates our use of domain-expert surveys to inform framework design rather than applying generic usability heuristics directly. \citet{springmeyer1992} conducted one of the earliest observational studies of how scientists use visualization tools, finding that data exploration is fundamentally iterative and that tools must support hypothesis refinement rather than just hypothesis testing. This finding resonates with our Dimension~2 (Interaction \& Customization), which evaluates whether models support diverse interaction paradigms including iterative retraining.
 
In the GIS-specific literature, \citet{HaklayTobon2003} proposed a framework for evaluating GIS usability that distinguishes between system-level usability (installation, configuration) and task-level usability (analytical workflow support). Our Dimensions~1 and~2 parallel this distinction: Dimension~1 evaluates the system-level barrier to entry, while Dimension~2 evaluates the task-level interaction paradigms available once access is achieved. \citet{Roth2015, Roth2013} developed a taxonomy of interaction primitives for cartographic interfaces, emphasizing that different analytical goals require different interaction modalities. This principle informs our profile-based approach to Dimension~2, where we assess a checklist of paradigms rather than imposing a single hierarchy.
 
More recently, the human-AI interaction literature has produced design guidelines specifically for AI-powered systems. \citet{amershi2019guidelines} distilled 18 design guidelines for human-AI interaction from a large-scale study at Microsoft, emphasizing principles such as making clear what the system can do, supporting efficient correction, and conveying the consequences of user actions. \citet{yang2020re} identified unique design challenges for AI systems, including the difficulty of communicating model uncertainty and the need to calibrate user expectations about system capabilities. These challenges map directly onto our Dimension~3 (Trust \& Transparency), which evaluates whether GeoFMs provide the uncertainty communication and validation support that responsible scientific use requires.
 
The open-source software sustainability literature also informs our framework. \citet{Katz2019} argue that scientific software sustainability depends not only on code availability but on community governance, documentation, and long-term maintenance commitments. \citet{Barker2022FAIR} propose the FAIR for Research Software (FAIR4RS) principles, extending the FAIR data principles (Findable, Accessible, Interoperable, Reusable) to software artifacts. Our Dimension~5 (Scientific Permanence \& Reproducibility) operationalizes these concerns for the GeoFM context, distinguishing between code availability (which most models achieve) and genuine archival permanence (which almost none achieve).
 
No existing framework integrates these perspectives, spanning usability evaluation, GIS interaction design, human-AI interaction, and scientific software sustainability, into a unified assessment instrument for geospatial foundation models.
 
\section{Motivating Study of Expert Requirements}
\label{sec:survey}
 
Before developing our evaluation framework, we conducted a pilot expert elicitation survey to ground our dimensions in empirical user needs. We present this study as exploratory motivation, given the limited sample size.

\subsection{Method}

We administered a structured survey to researchers at an international conservation monitoring center focused on biodiversity assessment and policy support. All participants maintained active research programs involving geospatial data analysis. The survey was anonymous, voluntary, unpaid, and conducted with institutional ethics approval. Eleven researchers participated, spanning ecology, conservation, earth science, and agriculture. Seven rated themselves as advanced users (4--5/5) of desktop GIS (mean comfort 3.7/5), while programming comfort was bimodally distributed with five participants rated themselves at 5/5, but two rated themselves at 1/5, with the remainder in between (mean 3.7/5). Command-line comfort was the lowest-rated skill (mean 2.9/5). Experience with geospatial data ranged from under one year (one respondent) to over 10 years (four respondents), and most participants worked with geospatial data at least weekly.
 
The survey comprised four sections (1) background and technical experience, (2) current workflow challenges and pain points, (3) familiarity with and perspectives on AI foundation models, and (4) ideal tool requirements and validation preferences. Questions employed Likert scales (1--5 for comfort, 1--7 for difficulty and importance), multiple choice, and open-ended responses. The full survey instrument is provided in \ref{app:survey}.
 
\subsection{Findings}
 
Five themes emerged that informed our evaluation framework design.
 
\subsubsection{Data Access and Preparation Dominate Workflow Bottlenecks}
 
Participants consistently identified user-centered barriers, in particular data access and preparation, rather than model-centric limitations as their primary bottleneck. On the 1--7 difficulty scale, the three highest-rated barriers were downloading large datasets (mean 4.7), navigating data portals (mean 4.6), and finding the right dataset (mean 4.3). By contrast, visualizing intermediate results (mean 3.1) and converting between formats (mean 2.9) were rated substantially lower. Packaging work for reproducibility was also rated as a significant barrier (mean 4.1).
 
Open-ended responses reinforced this pattern. One ecology researcher noted that ``downloading and preparing data is a huge bottleneck, especially obtaining time series of uncloudy satellite images for large number of points.'' Another described ``accessing and processing large volumes of data'' and the cascading problem of ``replicating upstream workflows to integrate new data when I realise something needs to change or be accounted for.'' A third respondent identified lack of reliable global data accepted by national experts as their primary obstacle, highlighting a data quality rather than data access concern.
 
Notably, when asked to describe their ideal tool, participants converged on data assembly functionality: ``a data collector/preprocessor that collects data from different sources (satellite, climate etc.) and downloads/preprocesses it for a list of points or a bounding box (longitude/latitude).'' Another wanted ``a tool that effectively directs you to the resources (data, map, literature) you need and helps you access/process them more easily.'' This desired functionality aligns closely with what GeoFM embeddings already provide, namely the pre-processed, multi-source representations for any location, suggesting a communication gap rather than a capability gap.
 
When asked about barriers to customizing an AI model, data preparation was again the most-cited barrier. 4 of 11 participants selected ``the effort of preparing my own data for the model,'' followed by training time (3/11). Only one participant cited computational cost as the primary barrier, despite computational requirements being a common concern in the GeoFM literature. This suggests that the actual bottleneck for domain experts is upstream of the model, in data curation and preparation, rather than in model execution.
 
\subsubsection{Awareness of Foundation Models Gap}
 
Familiarity with GeoFMs was distributed across a wide spectrum. Four participants reported they could ``use a pretrained model for a new task,'' three could ``explain the core concepts and methodology, but not implement it,'' three had only ``seen the term and have a general idea,'' and one could ``train or implement one from scratch.'' This distribution reveals a vast middle ground of semi-aware potential users who understand the concept but lack implementation knowledge.
 
Among the few who had directly used foundation models, experiences were mixed. One researcher had tested Tessera embeddings and ``easily and successfully generated representations (but now need to interpret!),'' highlighting the gap between embedding generation and downstream understanding. Another had worked with SpectralGPT, finding it ``performs better than a classic CNN'' but noting its limitation to single images rather than time series. A third had used Google AlphaEarth Embeddings, finding them serviceable but noting that ``identifying high-quality training data that doesn't already rely on remote sensing (and therefore may have been thrown into training data) is really hard.'' This is a data provenance concern rarely addressed in GeoFM documentation.

Preferences on the deployment format of models revealed a split that tracked with technical comfort. Of the four respondents who preferred full platform integration (e.g. Google Earth Engine, QGIS), all had programming comfort of 4 or below. Three respondents preferred pre-computed data files (embeddings), all with programming comfort of 4--5. Two preferred source code on GitHub. This pattern supports a tiered access model where technical users want embeddings as preprocessing steps in their analytical pipelines, while less technical users want AI capabilities seamlessly integrated into familiar interfaces.

\subsubsection{Trust Requires Uncertainty Calibration, Not Benchmarks}

When asked what would be most important for trusting AI outputs for scientific publication or policy recommendations, 5 of 10 respondents who answered prioritized ``clear visualization of where the model is confident and where it is uncertain.'' Three prioritized high accuracy on benchmark datasets, one prioritized alignment with domain knowledge, and one prioritized explanations of model reasoning for specific areas. This preference for uncertainty visualization held across technical comfort levels. Even the participant who could implement foundation models from scratch selected it.
 
For validation, participants relied primarily on qualitative, experience-grounded methods like visual spot-checking against high-resolution imagery (10/11), comparison with field data (9/11), and evaluating whether spatial patterns ``make ecological/geographical sense'' (9/11). Testing on known challenging areas was used by 7 of 11. Comparison with published benchmark statistics was also used by 8 of 11, but notably ranked below all qualitative methods except challenge-area testing. This suggests domain experts treat benchmarks as one input among several rather than as the primary arbiter of model quality, in contrast to the model-centric emphasis on benchmark leaderboards.
 
When rating model properties for importance on a 1--7 scale, participants most valued temporal pattern understanding (mean 6.5), open-source availability (mean 6.3), and state-of-the-art benchmark performance (mean 6.0), followed by physical law consistency (mean 5.7) and multi-scale analysis (mean 5.1). The high rating for open source (6.3) is notable: it suggests that scientific independence and reproducibility are near-top-of-mind concerns, consistent with our Dimension~5 rationale. The high rating for benchmarks (6.0) may seem to contradict the trust finding, but the distinction is important as participants value benchmarks as one property among several when \emph{evaluating} a model, while preferring uncertainty visualization as the most important feature for \emph{trusting} outputs in practice.
 
Among the AI capabilities rated for potential usefulness, anomaly detection (``identifying anomalous changes over time'') scored highest (mean 6.4/7), followed by what-if scenario generation (mean 5.5), gap-filling (mean 5.2), and finding similar regions (mean 4.4). The strong preference for anomaly detection and scenario generation suggests that domain experts see the greatest value in capabilities that augment their temporal monitoring workflows rather than in static spatial analysis.
 
\subsubsection{Interpretability as Process Understanding}
 
Participants expressed concerns about ``black box'' model behavior, but their needs extended beyond standard explainable AI. Responses clustered around maintaining connection of GeoFM outputs to underlying processes such as ``not being able to link patterns to processes/mechanisms in ecology'' and ``losing understanding of what is actually happening so it makes it difficult to properly interpret my results.'' A conservation expert with over 10 years of experience articulated a nuanced concern about ``the risk of over-polishing a data-to-decision-making stream that runs the risk of bypassing sufficient checks with experts, data, and stakeholders by virtue of its sheer ease and rapidity.'' Another respondent focused on the practical consequence: AI outputs ``being used without proper validation.''
 
These concerns highlight how interpretability requirements in scientific contexts differ from consumer AI applications. Experts need to maintain oversight over decision-making processes, not just understand individual predictions. In this sense, some experts may anticipate that as these tools become more accessible, opacity combined with ease of use could erode the careful, multi-stakeholder validation processes that responsible environmental decision-making requires.

\subsubsection{Diversity of Downstream Users}
 
One respondent represented a user category rarely considered in GeoFM development, namely the policy and decision-making professional who consumes model outputs but does not interact with models directly. This participant (programming comfort 1/5, preferred platform integration) described themselves as ``a user of the final product, and keen to help develop useful products'' and ``a user who tries to deploy products into decision-making arenas.'' Their primary concern was not model accuracy or interpretability but institutional acceptance: ``that countries will reject them, even if they are obviously better than things they are using.''
 
While a single respondent cannot support strong claims, this perspective surfaces a dimension neither our framework nor the broader literature addresses in the gap between scientifically valid outputs and institutional uptake.

\subsection{Implications for Framework Design}

Table~\ref{tab:survey_to_framework} summarizes how each survey theme informed specific framework dimensions.

\begin{table}[htbp]
\centering
\caption{Mapping from survey findings to framework dimensions.}
\label{tab:survey_to_framework}
\begin{tabular}{ll}
\toprule
\textbf{Survey Finding} & \textbf{Dimension} \\
\midrule
Data access as primary bottleneck & D1: Access \& Deployment \\
Deployment format preferences & D1 (Level 3--5 distinctions) \\
Uncertainty over benchmarks for trust & D3: Trust \& Transparency \\
Open-source as near-top priority & D5: Scientific Permanence \\
Interpretability as process oversight & D3 (T2, T3) \\
Anomaly detection \& workflow integration & D2: Interaction \& Customization \\
\bottomrule
\end{tabular}
\end{table}

\section{Evaluation Framework}
\label{sec:framework}
 
As motivated by the survey findings, we propose a seven-dimension evaluation framework. Each dimension reflects established HCI theory and is operationalized through clearly defined levels or checklists. We employ two types of measurement, using ordinal scales where a meaningful hierarchy exists (Dimensions 1, 4, 6, 7), and profile-based checklists where the landscape is multidimensional and a single ranking would impose a false hierarchy (Dimensions 2, 3, 5). This mixed approach respects the principle that different scientific tasks demand different types of support. We note in advance that some dimensions may show little variance across today's models; where this occurs, we interpret it not as a framework limitation but as a field-level diagnostic, a distinction we develop in Section~\ref{res:summary}.
 
\subsection{Dimension 1: Access \& Deployment}
 
\textbf{Guiding question:} ``How much prior knowledge does a user need to get this model to complete a bare minimum downstream task (e.g. a land cover classification on a 5$\times$5~km patch using embeddings)?''
 
This dimension addresses the initial cost of engagement. It is directly informed by Norman's concept of the ``gulf of execution'' \citep{Norman1988}, the cognitive and procedural distance between a user's goal and the actions required by the system to achieve it. For a domain expert whose goal is ``classify land cover in my study area,'' the gulf of execution varies depending on whether the system requires them to clone a repository and resolve CUDA driver conflicts (wide gulf) or upload a shapefile through a web interface (narrow gulf). Each level in our scale corresponds to a progressively narrower gulf, with specific technical prerequisites removed at each step.
 
This dimension also operationalizes Nielsen's usability heuristic of ``match between system and the real world'' \citep{Nielsen1994}, which holds that systems should speak the users' language and follow real-world conventions rather than requiring users to adapt to system-oriented concepts. A GeoFM released as a PyTorch checkpoint speaks the language of ML engineers but the same model exposed through a QGIS plugin speaks the language of geospatial analysts. Our levels capture this progression from developer-oriented to geospatial-oriented interfaces.
 
The level definitions are designed to reflect meaningful transitions in required user expertise. The boundary between Level~0 and Level~1 marks the difference between needing to \emph{train} a model and being able to \emph{use} a pretrained one, a distinction that separates research labs from potential practitioners. The Level~2/3 boundary captures whether the user has to independently construct a working environment or can begin from a functioning demonstration. The Level~4/5 boundary distinguishes between users who can write code against an API and those who require visual interaction.
 
Six levels are defined:
 
\begin{itemize}
    \item \textbf{Level 0:} Access only to source code. No pretrained weights provided; the user must execute pretraining from scratch. Effectively unusable outside major research labs.
    \item \textbf{Level 1:} Pretrained model weights provided (e.g. a checkpoint file), but the user is responsible for creating the computational environment from scratch with minimal guidance.
    \item \textbf{Level 2:} Weights plus a containerized environment (e.g. Dockerfile) or package manifest (e.g. \texttt{requirements.txt}). The process is simplified but still requires command-line fluency.
    \item \textbf{Level 3:} A self-contained, executable notebook or script with sample data and pre-written code for a complete demonstration. The primary barrier shifts from system setup to code execution.
    \item \textbf{Level 4:} The model outputs are hosted by the provider and accessible through a documented API. This eliminates user-side hardware and ML setup but requires programming skills.
    \item \textbf{Level 5:} Capabilities exposed through a graphical user interface (GUI), whether a web application, desktop program, or plugin for existing scientific software (e.g. QGIS, ArcGIS). Allows non-programmatic interaction.
\end{itemize}
 
\subsection{Dimension 2: Interaction \& Customization}
 
\textbf{Guiding question:} ``What interaction paradigms does this model's public release explicitly support and document for a user?''

A usable system should grant users a sense of agency and control over how they engage with its capabilities. This dimension is grounded in Shneiderman's Eight Golden Rules of Interface Design \citep{Shneiderman2013}, most notably the principle of ``supporting an internal locus of control,'' which posits that users feel most empowered when they are the initiators of actions rather than passive recipients of system outputs. A model that only supports a fixed inference pipeline treats the user as an operator, but one that supports fine-tuning, prompting, embedding extraction, and iterative retraining treats the user as a collaborator with meaningful control over the analytical process.
 
We use a profile-based structure with a base level plus a checklist of paradigms rather than a single ordinal scale, because different scientific tasks call for different modes of control. A single hierarchy would rank, say, fine-tuning above embedding extraction even when a given workflow needs the latter \citep{Roth2015}. The base level separates ``Research-Only'' (Level~1) releases, oriented toward the model's developers, from ``Supported'' (Level~2) releases that document an inference workflow for external users, which is a distinction paralleling the developer-versus-user-experience tension in scientific software \citep{Katz2019}. In practice, many GeoFM releases provide training scripts that technically permit inference but neither document nor support it for outside users. The five advanced paradigms (I1--I5) span the interaction modes we observed across the corpus and the human-AI interaction literature \citep{amershi2019guidelines}. Prompt-based exploration (I2) and interactive retraining (I4), shown to lower expertise barriers elsewhere, remain rare here.

\textbf{Base Level of Interaction:}
\begin{itemize}
    \item \textbf{Level 1 (Unsupported / Research-Only):} Artifacts released for research replication only. Adapting the code for inference on new data is not an explicitly documented or supported workflow.
    \item \textbf{Level 2 (Supported / Usable):} The release provides a clear, documented workflow for running inference on new, user-provided data.
\end{itemize}
 
\textbf{Supported Advanced Paradigms} (check all that apply):
\begin{itemize}
    \item \textbf{(I1) Expert Fine-Tuning:} Documented code and guidance for fine-tuning on a new dataset.
    \item \textbf{(I2) Prompt-Based Exploration:} Interaction via textual, visual, or query-by-example prompting as a documented method of use.
    \item \textbf{(I3) On-Demand Embedding:} A documented workflow for retrieving vector embeddings for user-provided data.
    \item \textbf{(I4) Interactive retraining:} A human-in-the-loop workflow of label $\rightarrow$ train $\rightarrow$ classify $\rightarrow$ validate $\rightarrow$ re-label.
    \item \textbf{(I5) Field Data Integration:} Documented workflow for incorporating common ground-truth and validation data formats.
\end{itemize}
 
\subsection{Dimension 3: Trust \& Transparency}
 
\textbf{Guiding question:} ``What specific features does this model provide to help a user understand, validate, and trust its outputs?''
 
For any AI system deployed in a high-stakes domain like environmental science, trust is not a feature but a prerequisite for responsible use. This dimension is informed by the principles of Human-Centered AI (HCAI) articulated by \citet{Shneiderman2020}, which emphasize that reliable, safe, and trustworthy systems require more than technical accuracy. They require mechanisms that allow users to understand system behavior, identify failure modes, and maintain meaningful oversight. The FAccT community's emphasis on treating AI systems as fallible tools requiring interrogation, rather than infallible oracles, further motivates this dimension \citep{Selbst2019}.

We operationalize trust through four independently assessed attributes rather than a single score, because the literature identifies distinct components \citep{miller2019explanation, bonneau2014overview} and our survey found that experts' priorities diverge sharply from developer defaults.

The decision to decompose trust in this way was motivated by our survey finding that domain experts prioritize spatially-explicit confidence estimates (T3) above all other trust attributes, while the ML community's default trust-building practice is benchmark replication (T1). A single composite trust score would obscure this misalignment. By assessing each attribute independently, we can identify precisely where the field's investment in trust-building aligns or diverges from user needs.
 
The threshold for T3 is deliberately specific. To qualify, a project must provide executable code producing per-prediction or per-pixel uncertainty values, and the documentation must explicitly refer to this capability. This strict operationalization prevents inflated scores from models that mention uncertainty in passing but provide no usable mechanism for generating it.
 
\begin{itemize}
    \item \textbf{(T1) Replicable Benchmarks:} Code and data provided to replicate the primary performance metrics reported in the paper.
    \item \textbf{(T2) Explainability Methods:} Documented, built-in tools for post-hoc explanation (e.g. attention maps, SHAP values).
    \item \textbf{(T3) Uncertainty Quantification:} A documented mechanism for estimating prediction uncertainty, whether native probabilistic outputs or a documented post-hoc method producing per-prediction or per-pixel uncertainty values.
    \item \textbf{(T4) Interactive Validation:} A GUI or documented code facilitating qualitative, visual comparison of outputs against ground truth or basemaps.
\end{itemize}
 
\subsection{Dimension 4: Community \& Support}
 
\textbf{Guiding question:} ``If a user gets stuck, what resources are available to help them?''
 
Usability does not reside solely in the interface; it exists within a sociotechnical system that includes documentation, tutorials, forums, and peer communities. This dimension is grounded in theories from Computer-Supported Cooperative Work (CSCW) and situated learning. Active forums serve as learning environments, not just support channels \citep{Lave1991}.
 
The levels in this dimension reflect a progression from minimal, developer-oriented documentation to rich, user-centric educational ecosystems. Level~1 (paper only) represents the minimum viable documentation for academic publishing but provides no support for practical use. Level~2 (detailed README) is the current norm in open-source ML, but as the scientific computing literature documents \citep{Wilson2014}, READMEs written by developers for developers systematically fail to anticipate the knowledge gaps of users from different disciplinary backgrounds. The critical transition occurs at Level~4, where documentation shifts from explaining \emph{how the code works} to explaining \emph{how to accomplish domain-specific tasks}.

Five levels are defined:
 
\begin{itemize}
    \item \textbf{Level 1:} Documentation limited to the academic publication. Support limited to issues on a largely unmaintained repository.
    \item \textbf{Level 2:} A detailed README explaining installation and usage, written for an audience with the same technical background as the authors.
    \item \textbf{Level 3:} A dedicated documentation website (e.g. Read the Docs) with conceptual overview, API reference, and separate installation and usage sections.
    \item \textbf{Level 4:} Tutorials aimed at domain scientists, including blog posts, commented notebooks, or video walkthroughs using domain-specific examples and accessible language.
    \item \textbf{Level 5:} A professionally managed, active community channel (e.g. Discord, Slack) in addition to robust documentation, with evidence of ongoing maintenance.
\end{itemize}
 
\subsection{Dimension 5: Scientific Permanence \& Reproducibility}
 
\textbf{Guiding question:} ``What guarantees does a scientist have that their work built on this model will be reproducible, extensible, and not dependent on a single entity in the long term?''
 
For scientific users, long-term usability is inseparable from reproducibility and permanence. While not a traditional HCI metric, it is for this group a first-order usability concern, bearing directly on their ability to publish, build upon, and trust their work over time. A researcher who builds a multi-year study on a GeoFM that is subsequently deprecated, made private, or reorganized faces not just inconvenience but potential invalidation of their scientific contribution.
 
This dimension draws on the FAIR for Research Software (FAIR4RS) principles \citep{Barker2022FAIR}, which extend the widely adopted FAIR data principles (Findable, Accessible, Interoperable, Reusable) to software artifacts. It also reflects the broader scientific software sustainability literature \citep{Katz2019}, which argues that code availability is a necessary but not sufficient condition for reproducibility. Genuine sustainability requires community governance, versioning, archival, and maintenance commitments.
 
We separate the base level of openness from specific provider policies because these capture distinct aspects of long-term viability. The base level (O1--O3) determines the degree of community control over the model's future. A closed-source model is entirely dependent on the provider's continued support, an open-core model can in principle be maintained by the community, and a fully archived model with a DOI is permanently retrievable regardless of what happens to the original provider. The policy checklist (P-S, P-P) captures additional commitments that affect scientific workflows. Service stability guarantees matter for researchers building on hosted APIs, while data portability matters for researchers who need to export results in standard geospatial formats for integration with other tools. This structure acknowledges data stewardship and digital preservation as first-class usability concerns for scientific practice.
 
\textbf{Base Level of Openness:}
\begin{itemize}
    \item \textbf{(O1) Closed-Source:} Core model and source code are proprietary.
    \item \textbf{(O2) Open Core:} Core model and software released under an open-source license.
    \item \textbf{(O3) Fully Archived:} The entire stack (paper, model, code, training data or replication recipe) is peer-reviewed and deposited in a permanent archive with a persistent identifier (e.g. DOI on Zenodo).
\end{itemize}
 
\textbf{Supported Policies} (check all that apply):
\begin{itemize}
    \item \textbf{(P-S) Service Stability:} A formal, public commitment to long-term support via a versioned API or deprecation schedule.
    \item \textbf{(P-P) Data Portability:} An explicit feature for exporting results in standard open formats (GeoTIFF, shapefile, NetCDF).
\end{itemize}
 
\subsection{Dimension 6: Multilingual Support}
 
\textbf{Guiding question:} ``To what extent can users access documentation and interfaces in multiple languages?''

A usable system should account for the linguistic diversity of its potential users \citep{Shneiderman2000}. For geospatial foundation models this is not just an inclusivity concern but a value-alignment one. The regions where these models offer the greatest marginal scientific value i.e. data-sparse areas with sparse existing labels, limited remote sensing infrastructure, and underdeveloped ground-truth networks, are disproportionately in lower- and middle-income countries and non-English-dominant research communities. The Earth observation capacity literature has long documented this gap. Uptake of EO has been slow and unevenly adopted across countries \citep{kavvada2020towards}, and the benefits of even openly accessible data accrue mainly to regions that already possess the infrastructure, compute, and trained workforce to absorb them \citep{wilson2025sdg}. A model's capacity to generalize from sparse supervision matters most where English-only documentation and an assumed familiarity with the ML frontier most exclude the local scientists best positioned to apply it. Multilingual access is therefore spatially correlated with where the technology's potential is highest, from deforestation monitoring in the Amazon to agricultural assessment across South and Southeast Asia, yet, as we find here, GeoFM documentation is overwhelmingly English-only.

\begin{itemize}
    \item \textbf{Level 0 (Monolingual):} All documentation and interfaces in one language only.
    \item \textbf{Level 1 (Limited Multilingual):} Documentation or interfaces in 1--2 additional maintained languages.
    \item \textbf{Level 2 (Broad Multilingual):} Documentation or interfaces in 3+ maintained languages.
\end{itemize}
 
\subsection{Dimension 7: Offline Usability}
 
\textbf{Guiding question:} ``Can users sustain field-relevant workflows under low-connectivity or offline conditions?''
 
Accessibility extends beyond software functionality to the physical and infrastructural contexts in which tools are deployed. This dimension aligns with the principle of contextual design articulated by \citet{Beyer1998}, which emphasizes understanding users' actual work environments to create tools that are effective in their real-world contexts, not just in laboratory or office settings. Assuming high-bandwidth, always-on internet access may not reflect the conditions under which many geospatial scientists work because ecological field stations often have intermittent connectivity, researchers in developing nations may face bandwidth constraints, and disaster response scenarios may involve degraded infrastructure.
 
An important nuance is that ``offline support'' for a foundation model should not mean running multi-billion-parameter inference on a laptop in a remote field station. The compute requirements of GeoFMs make local inference impractical for most domain experts regardless of connectivity. Rather, the relevant offline workflow involves precomputing embeddings for a region of interest before fieldwork, caching those embeddings locally, and then performing lightweight downstream analysis (classification, anomaly flagging, active learning label prioritization) on modest hardware while disconnected. In this workflow, the expensive inference happens once, in advance, on capable infrastructure, then the field scientist works with the resulting embeddings offline and synchronizes new labels or results upon reconnecting. This pre-cache-and-analyze pattern is particularly relevant for iterative fieldwork scenarios such as adaptive sampling, where embedding-based similarity search could guide which sites to visit next based on labels collected earlier in the same trip.

We note that this dimension is currently binary (online-only versus some offline support) and that a more granular scale distinguishing between incidental offline capability (the code happens to run locally) and deliberate field-deployment design (pre-caching, deferred sync, edge-optimized downstream models) may become appropriate as the ecosystem matures.

\begin{itemize}
    \item \textbf{Level 0 (Online-Only):} Core functionality requires stable, high-bandwidth internet.
    \item \textbf{Level 1 (Offline Support):} Meaningful tasks can be carried out under limited connectivity.
\end{itemize}

\section{Rating Methodology}
\label{sec:method}
 
\subsection{Model Corpus}

Our evaluation corpus comprises 89 geospatial foundation models. We included a model if it (i) was described as a foundation or pretraining model for Earth observation or remote sensing imagery, (ii) appeared in at least one of our source surveys \citep{feng2025tessera, zhu2024foundationsearthclimatefoundation, lu2025vision, xiao2025foundationmodelsremotesensing}, and (iii) had a publicly identifiable paper or release as of our evaluation window. We did not apply citation, venue, or performance thresholds, as our aim is a landscape snapshot. We note explicitly that this corpus is bounded by our source surveys and an October 2025--January 2026 evaluation window. Models released or updated after this window, or absent from all four source surveys, are necessarily excluded. The rapidly growing GeoFM literature means any fixed corpus is a snapshot, and we make no claim to exhaustiveness. Our findings describe field-level norms across a representative 89-model sample, which we expect to remain stable even as individual models are added. The corpus spans models released between 2021 and 2025, covering diverse architectures (vision transformers, contrastive learning, masked autoencoders), modalities (optical, SAR, multispectral, hyperspectral), and scales (regional to global).

\subsection{Rating Protocol}
 
Two raters independently evaluated each model in the corpus between October 2025 and January 2026, based on publicly available artifacts at the time of evaluation. For each model, raters examined the published paper, the code repository (if available), associated documentation and websites, and any hosted interfaces or APIs. Ratings were recorded using a standardized template. For Dimension~1, if a model received Level~0 (no usable access), the remaining dimensions were not rated, as the model is effectively inaccessible for evaluation. As the GeoFM landscape evolves rapidly, some models may have updated their public releases since our evaluation window, and as such our results represent a systematic snapshot rather than a continuously updated index.
 
The raters first independently coded a calibration sample of 20 models drawn from across the accessibility spectrum. After computing inter-rater reliability on this overlap set (Section~\ref{sec:irr}), the remaining models were divided between the two raters. Each rater coded approximately 30 additional unique models in addition to the 20-model overlap, yielding 89 unique models in total.

\subsection{Inter-Rater Reliability}
\label{sec:irr}

We computed inter-rater agreement on the 20-model calibration sample, yielding 244 pairwise rating comparisons across all dimensions and attributes. Overall observed agreement was 93.0\%. We report both nominal and weighted agreement statistics. For binary attributes, nominal Cohen's $\kappa = 0.92$ and Gwet's AC1 $= 0.95$. For ordinal dimensions, quadratic-weighted $\kappa = 0.95$ and AC2 $= 0.95$, all indicating excellent agreement. We report AC1 alongside $\kappa$ because several attributes exhibited extreme prevalence skew (e.g. nearly all models scored identically on T1, T3, D6, D7), which inflates $\kappa$'s chance-agreement baseline and produces paradoxically low $\kappa$ values despite observed agreement. This is a well documented phenomenon \citep{Gwet2008, FeinsteinCicchetti1990}, and AC1 is robust to prevalence effects and better reflects true agreement under these conditions.
 
The primary source of disagreement was the Level~2/3 boundary in Dimension~1 (Access \& Deployment), where one rater systematically rated models with both a package manifest and a demonstration notebook as Level~3, while the other rated some of these as Level~2. This was resolved through discussion and clarification of the boundary criterion before the full coding proceeded. Per-dimension agreement statistics are provided in \ref{app:irr}.
 
\section{Results}
\label{sec:results}
 
We report results organized by dimension, focusing on both discriminative findings (dimensions that differentiate models) and diagnostic findings (dimensions that reveal systemic field-level gaps).
 
\subsection{Dimension 1: Access \& Deployment}
 
The distribution of access levels across 89 models is summarized in Table~\ref{tab:d1_results}. Nearly 29\% of models ($n=26$) scored Level~0, meaning that no pretrained weights are publicly available and the model cannot be used without re-executing pretraining. An additional 22\% ($n=19$) scored Level~1, providing weights but minimal environment guidance. Together, over half of the surveyed GeoFMs (50.6\%) require users to either pre-train from scratch or build a complex computational environment with little support.
 
\begin{table}[htbp]
\centering
\caption{Distribution of Access \& Deployment levels across 89 GeoFMs.}
\label{tab:d1_results}
\begin{tabular}{clcc}
\toprule
\textbf{Level} & \textbf{Description} & \textbf{Count} & \textbf{\%} \\
\midrule
0 & Source code only (no weights) & 26 & 29.2 \\
1 & Weights, minimal guidance & 19 & 21.3 \\
2 & Weights + environment config & 18 & 20.2 \\
3 & Notebook/script with demo & 20 & 22.5 \\
4 & Hosted API & 1 & 1.1 \\
5 & Graphical user interface & 5 & 5.6 \\
\bottomrule
\end{tabular}
\end{table}
 
\begin{figure}[htbp]
\centering
\includegraphics[width=\columnwidth]{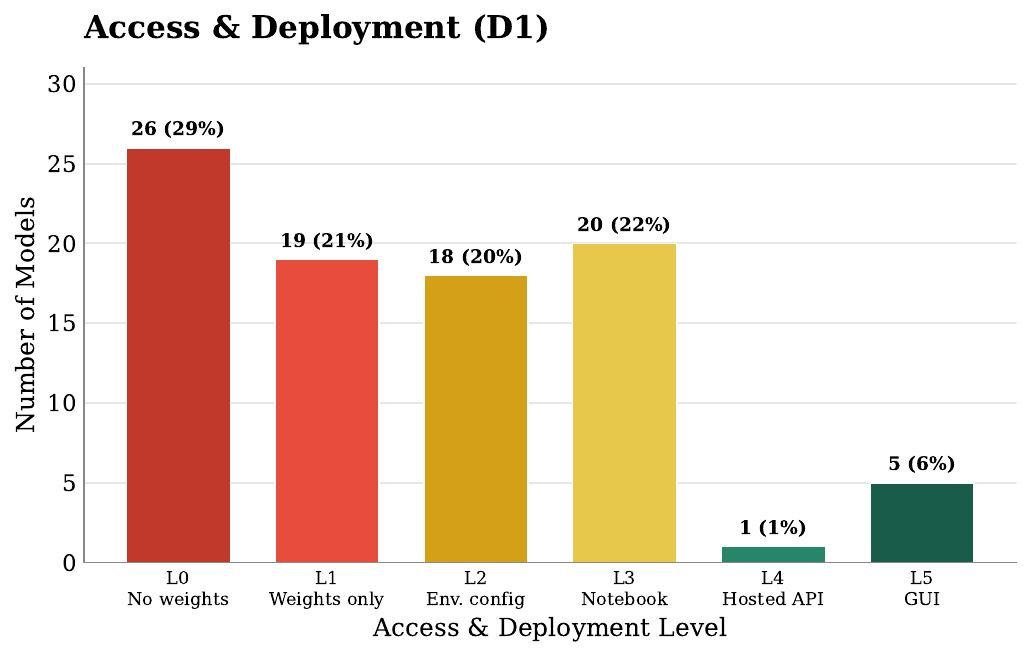}
\caption{Distribution of Access \& Deployment levels (Dimension~1) across 89 geospatial foundation models.}
\label{fig:d1}
\end{figure}

At the other end, only six models (6.7\%) provide access at Level~4 or above, through a hosted API or GUI. These are MOSAIKS, Google's AlphaEarth, Prithvi (via the IBM/NASA TerraTorch ecosystem), AIEarth, SatLas, and Tessera. The distribution, visualized in Fig.~\ref{fig:d1}, reveals a gap. Level~3 (notebooks with sample data) represents the current ceiling for the vast majority of open-source GeoFMs, while accessible interfaces remain the exception.

The 26 Level~0 models were not rated on subsequent dimensions, as they are effectively inaccessible. All subsequent results are reported for the models rated Level~1 or above.

\subsection{Dimension 2: Interaction \& Customization}

\begin{figure*}[htbp]
\centering
\includegraphics[width=\textwidth]{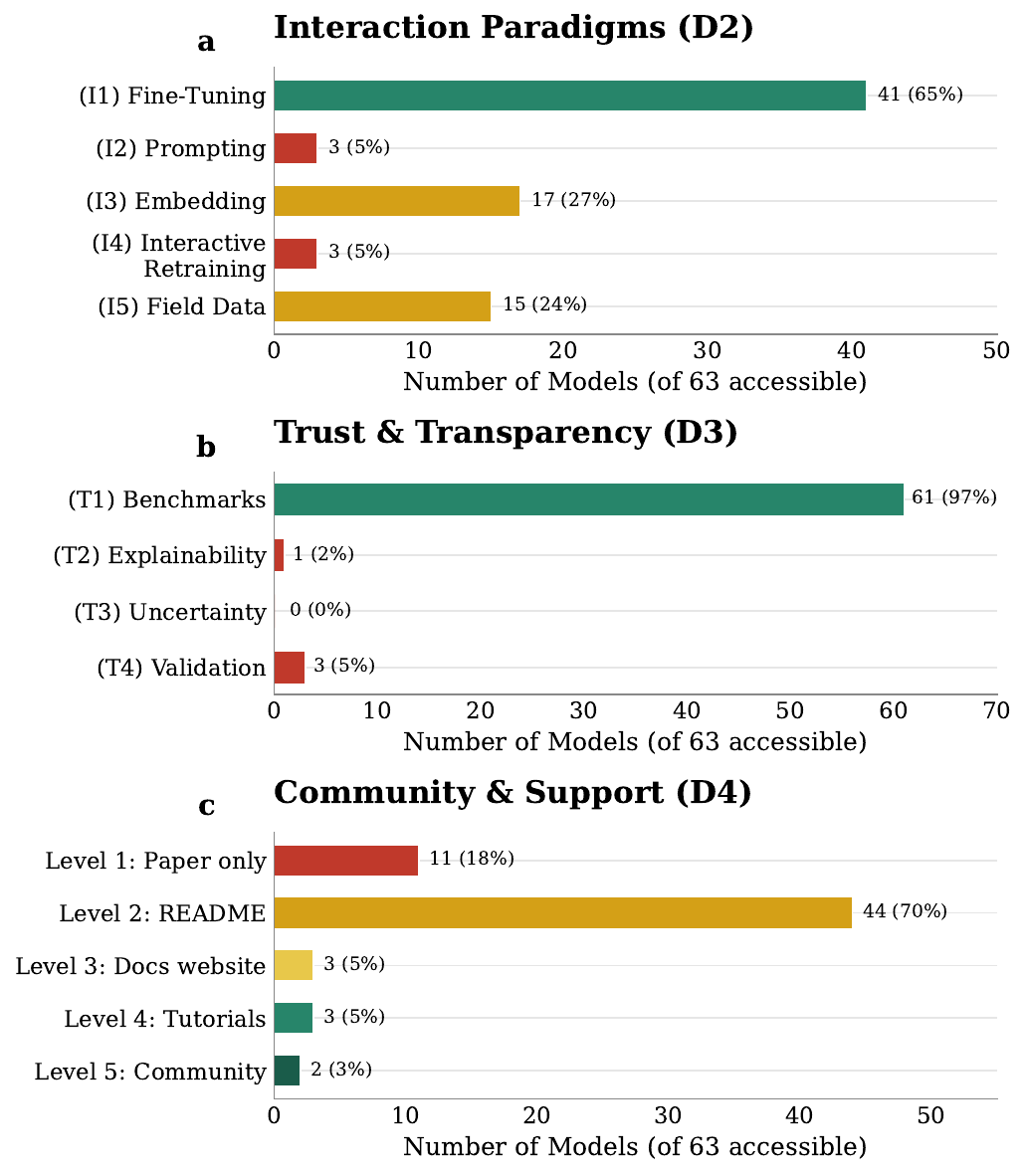}
\caption{Distribution of \textbf{(a)} supported interaction paradigms (Dimension~2), \textbf{(b)} trust and transparency attributes (Dimension~3), and \textbf{(c)} community and support levels (Dimension~4) across 63 accessible GeoFMs.}
\label{fig:d234}
\end{figure*}

Of the 63 accessible models, 52 (85.2\%) provide a supported workflow for user-provided inference (Base Level~2), while 9 (14.8\%) release artifacts for research replication only (Base Level~1). The distribution of supported advanced paradigms is shown in Fig.~\ref{fig:d234}a.

Fine-tuning support (I1, 65\%) is the dominant paradigm, consistent with the field's emphasis on transfer learning as the primary interaction mode. However, this paradigm requires the most ML expertise from users. Prompt-based exploration (I2) and interactive retraining (I4), which would lower the expertise barrier, are each supported by three models (5\%). On-demand embedding generation (I3, 27\%) is supported by roughly one-quarter of models, despite our survey finding that domain experts prefer precomputed embeddings as their access format.

\subsection{Dimension 3: Trust \& Transparency}
 
 
The trust dimension reveals the largest field-level gap (Fig.~\ref{fig:d234}b). Nearly all models (96.8\%) provide replicable benchmarks, reflecting academic norms around reproducibility of reported metrics. Beyond this baseline, the trust infrastructure is almost entirely absent. Only one model provides built-in explainability methods. Zero models provide documented uncertainty quantification, the feature that our survey found domain experts prioritize most highly for establishing trust. Only two models support interactive validation workflows.

This finding represents evidence of misalignment between GeoFM development and user needs. The field has invested heavily in benchmark replication, while providing none of the spatially-explicit uncertainty estimates that our survey respondents most often prioritized, which is a mismatch that we suggest is worth investigating at larger scale.

This gap persists even at the top of the accessibility spectrum. For example, Tessera and SatLas both add interactive validation (T4) on top of replicable benchmarks, yet neither provides documented uncertainty quantification. The most paradigm-complete releases in the corpus reach the ceiling of available trust infrastructure and still lack the one attribute our survey respondents prioritized most.

\subsection{Dimension 4: Community \& Support}

 
 
The distribution of community support levels is shown in Fig.~\ref{fig:d234}c. The overwhelming majority of models (70\%) provide a detailed README as their primary documentation, written for an audience with the same technical background as the authors. Only eight models (12.7\%) provide documentation at Level~3 or above, indicating dedicated documentation websites, domain-oriented tutorials, or active community channels. There are only two models (Tessera and Prithvi) achieving Level~5, with an active community ecosystem including documentation, tutorials, and community support.
 
\subsection{Dimension 5: Scientific Permanence \& Reproducibility}

Nearly all accessible models (96.8\%) are Open Core (O2), released under open-source licenses on GitHub. Only one model is Closed-Source (O1) and one is Fully Archived (O3) with a DOI and permanent deposit. Service stability guarantees (P-S) are provided by 7 models (11.1\%), and data portability features (P-P) by 8 models (12.7\%). The near-uniformity at O2 reflects the academic open-source norm but obscures the absence of long-term archival practices. GitHub repositories can be deleted, reorganized, or abandoned. Without permanent archiving, scientific reproducibility can remains fragile.

\subsection{Dimension 6: Multilingual Support}
 
Of 61 models with available data, 60 (95.2\%) provide documentation in one language only (Level~0). Three models (4.8\%) offer limited multilingual support (Level~1). No model achieves Level~2 (broad multilingual). This near-total absence of multilingual documentation represents a barrier for scientists in non-English-dominant research communities and conflicts with the global scope of Earth observation applications (all models assessed were described in either English or Chinese).

\subsection{Dimension 7: Offline Usability}
 
Of 63 accessible models, 61 (96.8\%) support offline workflows (Level~1), while 2 (3.2\%) require persistent internet connectivity (Level~0). However, this high prevalence of offline support is largely incidental. Most open-source models are downloaded and run locally, which technically permits offline use. No model in our corpus provides the deliberate field-deployment features described in our framework rationale, such as embedding pre-caching for a region of interest, edge-optimized downstream inference, or deferred synchronization of field-collected labels. The two online-only models are platform-locked systems requiring cloud connectivity. As discussed in Section~\ref{sec:framework}, the distinction between incidental and deliberate offline support may warrant a more granular scale in future iterations of the framework.

\subsection{Summary: Discriminators and Diagnostics}
\label{res:summary}

Table~\ref{tab:summary} summarizes the landscape across all dimensions, distinguishing dimensions that currently \emph{discriminate} between models from those that serve as \emph{field-level diagnostics}.
 
\begin{table*}[htbp]
\centering
\caption{Summary of evaluation results across 89 GeoFMs. Discriminative dimensions differentiate individual models; diagnostic dimensions reveal systemic field-level patterns.}
\label{tab:summary}
\hspace*{-2cm}
\begin{tabular}{llll}
\toprule
\textbf{Dimension} & \textbf{Function} & \textbf{Key Finding} & \textbf{Variance} \\
\midrule
D1: Access \& Deployment & Discriminative & 51\% at Level 0--1; only 6\% at Level 4--5 & High \\
D2: Interaction paradigms & Discriminative & I1 (65\%) dominates; I2, I4 near zero & Moderate \\
D3: Trust \& Transparency & Diagnostic & 0\% uncertainty; 97\% benchmarks only & Very low \\
D4: Community \& Support & Discriminative & 71\% rely on README alone & Moderate \\
D5: Permanence & Diagnostic & 97\% Open Core; 2\% archived with DOI & Very low \\
D6: Multilingual & Diagnostic & 95\% monolingual & Very low \\
D7: Offline & Diagnostic & 97\% offline-capable by default & Very low \\
\bottomrule
\end{tabular}
\end{table*}
 
Dimensions 1, 2, and 4 currently function as \emph{discriminators}, where they rank and differentiate individual models, helping practitioners identify which systems best meet their needs. Dimensions 3, 5, 6, and 7 currently function as \emph{field-level diagnostics}, as their low variance reveals where the geospatial foundation model community has collectively converged on a narrow band of practices and where systemic investment is needed. Both types of findings are core contributions of a systematization of knowledge. The first helps practitioners choose between models today, the second tells the community where to invest effort going forward. As the ecosystem matures, diagnostic dimensions should begin showing greater variance and transition toward discriminative utility.
 
\subsection{The User's Progression From Checkpoint to Platform}
 
Aggregate statistics convey the landscape but not the experience. To make the accessibility levels concrete, we trace what a hypothetical domain expert, a conservation scientist seeking to generate land cover embeddings for a study area, would encounter at each stage of the accessibility spectrum. At each level, we use a representative model from our corpus to illustrate the typical user experience. These models are not singled out as exemplary or deficient, as they only reflect the standard practice at its level of accessibility and represents a common, understandable design choice with distinct trade-offs.

\subsubsection{Level 0: No Usable Access}
 
Before our hypothetical user even reaches a usable starting point, they contend with the most common outcome in the corpus, where no pretrained weights are available at all. 26 models (29.2\%) in our survey fall into this category. The user finds a published paper and, in some cases, a repository containing model architecture definitions and training scripts, but no checkpoint files. To use the model, the user would need to first acquire or construct a large-scale pretraining dataset (often terabytes of satellite imagery), configure and execute a computationally intensive pretraining pipeline, and only then begin the downstream task they actually cared about. This is not a usability barrier in the conventional sense, as it is more of a prerequisite that places the model entirely outside the reach of anyone without access to a well-equipped research computing facility. These 26 models were not rated on subsequent dimensions in our evaluation, as there is no user experience to assess.

\subsubsection{Level 1: Weights with Minimal Guidance}

At the lowest level of usable access, our user may find a GitHub repository containing a model checkpoint and training scripts. Consider DINO-MM \citep{wang2022dinomm}, a multimodal self-supervised model for remote sensing released in 2022. The repository provides pretrained checkpoint files and code, but no \texttt{requirements.txt}, no Dockerfile, and a minimal README. The user has to independently determine compatible versions of Pytorch, CUDA, and supporting libraries, locate or construct a compatible data loader for their satellite imagery, and write inference code to extract predictions or embeddings for their study area. None of these steps are documented. The interaction paradigm is research-only (Base Level~1). The code is released for replication of the paper's experiments, not for application to new data. No advanced paradigms (fine-tuning, embedding extraction, prompting) are documented. Community support consists of the paper itself and an issue tracker with limited activity. Trust infrastructure is limited to replicable benchmarks.
  
\subsubsection{Level 2: Environment Configuration Provided}
 
One step up, our user finds a repository that includes environment specification files alongside model weights and code. SatMAE \citep{Cong2022}, a masked autoencoder for temporal and multi-spectral satellite imagery, provides pretrained checkpoints and a YAML environment file for Conda-based setup. The user can create a compatible Python environment with a single command rather than manually resolving dependencies. However, the workflow remains command-line driven: fine-tuning requires running \texttt{main\_finetune.py} with appropriate configuration flags, and embedding extraction requires reading the source code to identify the correct forward pass. Documentation consists of a detailed README written for an audience with ML engineering fluency. Fine-tuning and embedding extraction are supported (I1, I3) but described in technical terms that assume familiarity with PyTorch training loops, data loaders, and checkpoint management.
 
This level eliminates the most frustrating barrier of Level~1, dependency resolution, but the expertise requirements remain substantial. The user must be comfortable with command-line tools, Python scripting, and ML concepts. Eighteen models (20.7\%) in our corpus occupy this level.

\subsubsection{Level 3: Notebooks and Demonstrations}
 
At Level~3, the primary barrier shifts from environment setup to code execution. Clay \citep{ClayModel}, a foundation model for Earth observation, is an example of this level. The project provides a dedicated documentation website, environment specification files, and, critically, tutorial notebooks that walk users through complete workflows including fine-tuning, embedding generation, and interactive retraining. Youtube video tutorials supplement the written documentation, explaining use cases in accessible language. The user can follow a notebook from data loading to embedding extraction without writing code from scratch.
 
Clay achieves the highest interaction paradigm coverage in our open-source corpus. Fine-tuning (I1), on-demand embedding (I3), and interactive retraining (I4) are all documented through tutorials. Community support reaches Level~4, with domain-oriented educational materials rather than developer-oriented READMEs. The project maintains an active web presence with evidence of ongoing development and a service stability commitment (P-S).

This level represents the best-case scenario for open-source GeoFMs that do not provide hosted services. Twenty models (23.0\%) achieve this level, though few match Clay's breadth of documented paradigms. Even at Level~3, the user must still install Python, manage a local environment, and execute notebooks, a workflow accessible to computationally literate scientists but not to GIS analysts accustomed to graphical interfaces.

\subsubsection{Level 5: Platform Integration}
 
At the top of the accessibility spectrum, our user encounters models accessible through graphical interfaces that require no programming. Only five models in our corpus achieve this level, and they illustrate distinct approaches with fundamentally different trade-offs.

\textbf{The open ecosystem approach.} Prithvi \citep{jakubik2023foundationmodelsgeneralistgeospatial}, developed by IBM and NASA, achieves Level~5 through interactive HuggingFace Spaces demonstrations while also providing the full open-source stack for programmatic use. It is the only model in our corpus to achieve Level~5 on Community \& Support, with an active discussion forum, domain-oriented tutorials, and the TerraTorch library for fine-tuning. It supports three advanced interaction paradigms (I1, I3, I5) and provides both service stability guarantees and data portability in standard geospatial formats. However, even Prithvi, the most complete ecosystem in our survey, provides no uncertainty quantification (T3), no explainability methods (T2), and no interactive validation tools (T4). Its trust infrastructure, like that of every other model in the corpus, is limited to replicable benchmarks.

\textbf{The open embedding-as-data approach.} Tessera \citep{feng2025tessera, feng2026tesserav2} achieves Level~5 through an interactive web map and a dedicated Python library for accessing precomputed embeddings. It is the only model in our corpus to support all five interaction paradigms (I1--I5) and one of only two to achieve Level~5 on Community \& Support, with an active Zulip chat, tutorials, and a maintained website. Tessera distributes precomputed embeddings as downloadable data products, combining the accessibility of a platform with the scientific independence of open-source code and data.

\textbf{The platform-locked approach.} Google AlphaEarth \citep{brown2025alphaearthfoundationsembeddingfield} achieves Level~5 through full integration with Google Earth Engine, providing a GUI, an API, documentation, and video tutorials. For accessibility, it represents the gold standard. A user can explore embeddings, visualize results, and export data without leaving the Earth Engine interface. It is the only model in our corpus to provide pre-computed embeddings at global scale as a ready-to-use data product.

The tradeoff is scientific independence. AlphaEarth is the only Level~5 model rated Closed-Source (O1) on Dimension~5: the model architecture, training code, and training data are proprietary. Researchers cannot inspect what the embeddings represent, cannot validate whether representations capture scientifically relevant phenomena rather than visual correlations, and cannot reproduce the embedding generation process. The system requires persistent internet connectivity (Level~0 on Dimension~7), and continued access depends entirely on Google's service commitments. For prototyping and exploratory analysis, this trade-off may be acceptable. For long-term scientific programs that require reproducibility and auditability, it introduces dependencies that some researchers would consider unacceptable.
 
\subsubsection{Summary of Archetypes}
 
Table~\ref{tab:archetypes} summarizes the profiles of these representative models, illustrating how the framework dimensions interact to produce qualitatively different user experiences.
 
\begin{table*}[htbp]
\centering
\caption{Profiles of representative GeoFMs at different accessibility levels, illustrating trade-offs across framework dimensions.}
\label{tab:archetypes}
\hspace*{-1.5cm}
\begin{tabular}{lccccccc}
\toprule
\textbf{Model} & \textbf{D1} & \textbf{D2 Base} & \textbf{Paradigms} & \textbf{D3 (T1--T4)} & \textbf{D4} & \textbf{D5} & \textbf{D7} \\
\midrule
DINO-MM (2022) & 1 & 1 & 0/5 & T1 only & 1 & O2 & 1 \\
SatMAE (2022) & 2 & 2 & I1, I3 & T1 only & 2 & O2 & 1 \\
Clay (2024) & 3 & 2 & I1, I3, I4 & T1 only & O2, P-S & 4 & 1 \\
Prithvi (2023) & 5 & 2 & I1, I3, I5 & T1 only & O2, P-S, P-P & 5 & 1 \\
AlphaEarth (2025) & 5 & 2 & I3 & T1 only & O1, P-S, P-P & 4 & 0 \\
Tessera (2025) & 5 & 2 & I1--I5 & T1, T4 & O2, P-S, P-P & 5 & 1 \\
\bottomrule
\end{tabular}
\end{table*}
 
The archetype comparison reveals a pattern. As access and community support improve across models, the trust infrastructure remains uniformly thin. The column for Dimension~3 reads ``T1 only'' for every model except for Tessera regardless of accessibility level (Table~\ref{tab:archetypes}). One of the most accessible models in our corpus and the least accessible share identical trust profiles. This suggests that the trust gap identified in our aggregate results is not confined to under-resourced releases.

\section{Discussion}
\label{sec:discussion}
 
We suggest that the results reveal three interlocking problems: an accessibility cliff that excludes most domain experts, a trust infrastructure that is oblivious to what those experts need most, and a distribution model that treats code release as the endpoint rather than the beginning of accessibility. Together, these reflect a field that has optimized along model-centric dimensions while leaving user-centered ones largely unaddressed. In this section, we move beyond documenting these gaps to discuss what the model developing community can do about them.
 
\subsection{Bridging the Missing Middle}
 
The sharp dropoff between Level~3 and Level~4 in Fig.~\ref{fig:d1} defines the central design challenge. Most open-source GeoFMs top out at notebook-level access, while the few models offering GUI-based interaction are either backed by major institutional investment (Prithvi, AlphaEarth) or commercial platforms (AIEarth). This creates the impression that accessible interfaces require resources beyond the reach of typical research groups. We argue this is no longer true.

The broader ML community has developed lightweight tools that dramatically lower the cost of building accessible interfaces. HuggingFace Spaces, Gradio, and Streamlit allow researchers to wrap existing Python inference code in web-based interfaces with minimal engineering effort. In many cases, a Level~2 model can be elevated to Level~5 in days rather than months. The GeoFM community has largely not adopted these tools despite their availability. We suggest this is primarily an incentive problem, as academic publishing rewards benchmark improvements and architectural innovations, not interface development. A community norm that treats a Gradio demo or a HuggingFace Space as a standard component of a model release, alongside the paper, the weights, and the training code, would shift many models from Level~1--2 to Level~4--5 at negligible cost.

For models where interactive demos are insufficient, the most impactful intervention is the distribution of pre-computed embeddings as structured data products. The operational template for this already exists. Take, for example, NASA's Earthdata program distributes derived data products at planetary scale, the Copernicus Climate Data Store serves terabytes through a versioned API, and Google Earth Engine hosts analysis-ready data for millions of users. The missing piece is that almost nobody has applied this template to foundation model embeddings. At the time of our evaluation (October 2025--January 2026), only one open model in our corpus (Tessera \citep{feng2025tessera}) distributed pre-computed embeddings, and one closed platform (AlphaEarth \citep{google_satellite_embedding_v1_annual}) offered them within its ecosystem. Early indications suggest this is beginning to change \citep{wang2025copernicus, ye2026modelplacetimeremote}, but the practice remains far from standard.

A global inference pass is costly but bounded, one-time per model version, and centralizable. Contrast this with the current pattern, in which every research group independently downloads the same weights and runs overlapping inference on their own hardware, duplicating computational expense across hundreds of institutions. Centralized pre-computation, funded as public infrastructure through space agency programs or climate research initiatives such as the EU's Destination Earth, would be more efficient. The result would be versioned embedding data products with clear provenance, served through APIs, that domain experts can query without ever touching model weights. This is precisely the access pattern our survey respondents described wanting.

Notably, Tessera demonstrates that this pattern is achievable outside the major-institution setting. Unlike AlphaEarth, it distributes precomputed embeddings as downloadable data products while keeping its code, weights, and embeddings open, showing that the embedding-as-data-product model does not inherently require either corporate infrastructure or the closed-source tradeoff. This is an existence demonstration that the access pattern our survey respondents described wanting is reachable by a conventional research group.

\subsection{Closing the Trust Gap}
 
The complete absence of uncertainty quantification across 63 accessible models represents the most direct evidence of misalignment between GeoFM development and user needs. Our survey found that domain experts rank spatially-explicit confidence estimates as their primary requirement for trusting AI outputs, yet the field has invested almost exclusively in replicable benchmarks, the trust attribute that experts rank lower for establishing trust in outputs.

We acknowledge that uncertainty quantification for foundation model embeddings presents significant technical challenges. High-dimensional embedding spaces resist straightforward probabilistic interpretation, and the separation between the foundation model (which produces embeddings) and the downstream task head (which produces predictions) complicates the question of where uncertainty should be estimated. However, the total absence of even well-established approaches such as ensemble disagreement or calibrated prediction intervals suggests the gap is one of priority rather than feasibility. Several of these methods are well-established in the broader deep learning literature and could be adapted for geospatial embeddings with research effort.

We propose that the GeoFM community should consider uncertainty quantification as a first-class component of model architecture, not an optional addition. Concretely, this means (1) downstream task heads should report calibrated prediction intervals, not just point estimates; (2) embedding-level anomaly detection (e.g. flagging inputs that fall far from the training distribution in embedding space) should be a standard feature; and (3) validation interfaces should support spatially-explicit uncertainty visualization, enabling the kind of ``geography of model ignorance'' that domain experts described wanting in our survey. These capabilities would allow scientists to direct field verification efforts toward high-uncertainty regions rather than high-probability detections, a fundamentally different and more efficient use of limited fieldwork resources.

\subsection{Toward Usability Metadata Standards}
 
Beyond infrastructure and tools, our findings suggest the need for community-level norms around usability reporting. Model cards, as popularized by \citet{Mitchell2019}, have become a standard mechanism for documenting model characteristics. We propose extending this practice to include structured usability metadata. A model release could declare its framework profile, for example ``D1: Level~2, D2: I1/I3, D3: T1 only, D4: Level~2, D5: O2,'' as part of its model card or README, using perhaps the vocabulary our framework provides, or other standards that the community comes to a consensus on.

This would serve two purposes. First, it would make usability dimensions visible and comparable, allowing practitioners to select models based on accessibility as well as benchmark performance. Second, it would create a feedback signal. If publication venues like RSE, TGRS, or CVPR began requiring a usability statement alongside existing reproducibility checklists, model developers would have an incentive to invest in the dimensions that currently score lowest. The framework we present here provides a ready-made vocabulary for such statements. This approach imposes no significant technical burden and requires only that developers assess and report what their release provides.
 
\subsection{Temporal Trends}

\begin{figure*}[htbp]
\centering
\includegraphics[width=\textwidth]{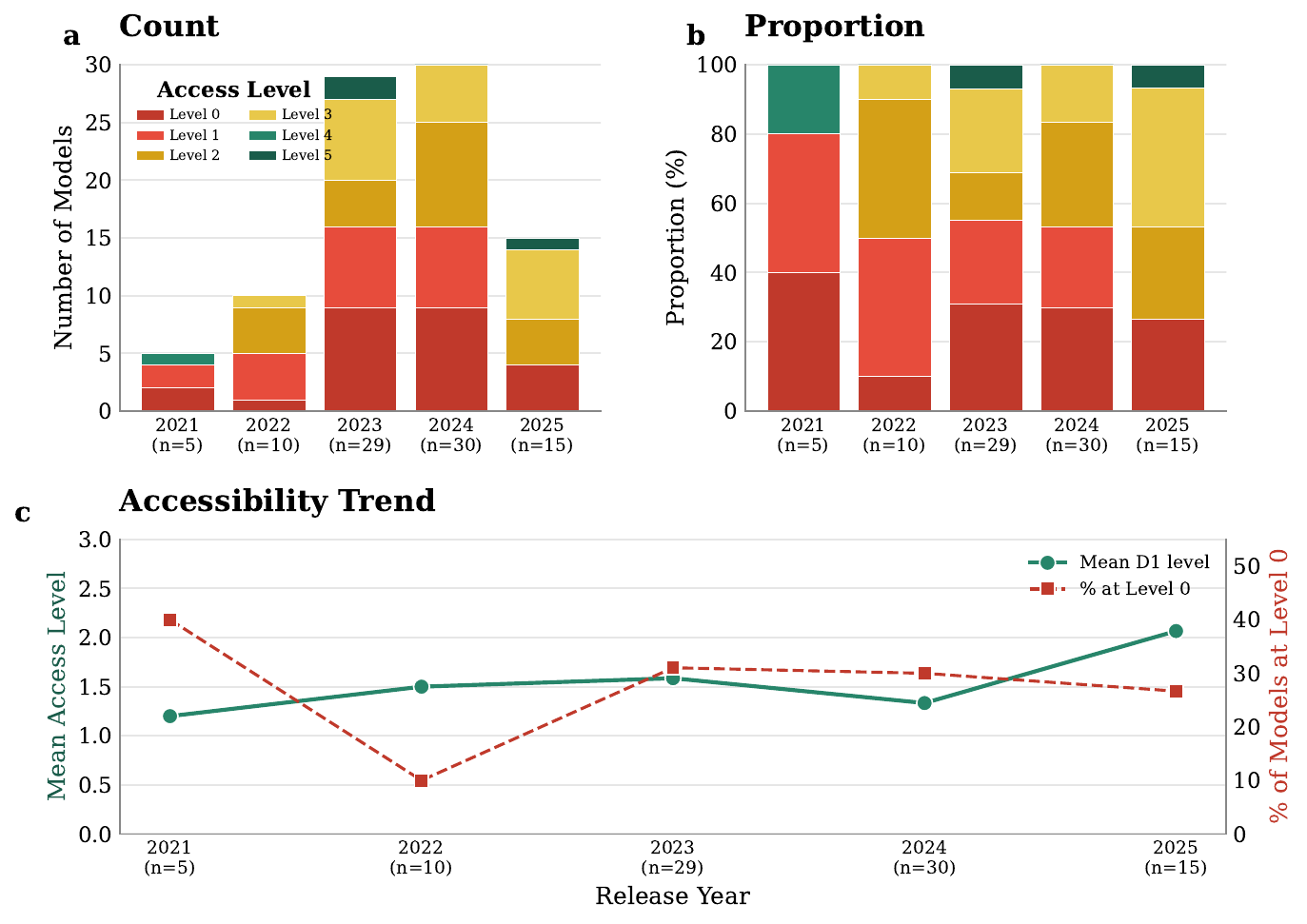}
\caption{Temporal trends in GeoFM accessibility. \textbf{(a)} Raw count of models 
by access level and release year. \textbf{(b)} Proportion of models at each access 
level, normalized by year, controlling for the uneven growth in model releases. 
\textbf{(c)} Mean access level (green) and percentage of models at Level~0 (red) 
over time. Sample sizes vary across years (shown below axis).}
\label{fig:temporal}
\end{figure*}

Because our corpus spans models released from 2021 to 2025, we can examine whether accessibility is improving over time. Fig.~\ref{fig:temporal} shows the distribution of Dimension~1 levels by release year. 

Sample sizes per year are small and uneven ($n = 5$ for 2021 to $n = 15$ for 2025), so the following should be read as suggestive. With that caveat, mean access level remained relatively flat from 2022 to 2024 (1.3--1.6), but rose to 2.2 for 2025 models, with 54\% of 2025 releases reaching Level~3 or above compared to 27\% in 2024. The proportion of Level~0 models (no usable access) declined from 34\% in 2023 to 23\% in 2025.
 
This trend is cautiously encouraging but should be interpreted carefully given the small and uneven sample sizes across years ($n=5$ for 2021, $n=13$ for 2025). The improvement in 2025 may reflect a shift in community norms, as several high profile 2025 releases included notebooks and demo scripts as standard practice, or it may reflect selection effects in our corpus. Regardless, the 2022--2024 plateau suggests that for the early GeoFM era, accessibility was not improving despite rapid growth in the number of models released. The improvement, where it exists, has been concentrated in Dimension~1 (access); the trust, permanence, and multilingual dimensions show no temporal improvement at all.
 
\subsection{Diagnostic Dimensions as Field-Level Signals}
 
Several dimensions produced very low variance, with nearly all models scoring identically. Rather than indicating framework limitations, these findings constitute field-level diagnostics that identify systemic underinvestment.
 
\textbf{Scientific Permanence (D5):} 96.8\% of models are Open Core on GitHub, but only 1.6\% are permanently archived with a DOI. This reflects the academic norm of ``open source on GitHub'' without the additional step of permanent, citable archival that the FAIR4RS principles \citep{Barker2022FAIR} call for. For long-term scientific reproducibility, this is insufficient as repositories can be deleted, reorganized, or made private, invalidating downstream work that depends on specific model versions. Depositing model weights and code in Zenodo or a similar archive is a low-effort, high-impact practice that the community should consider adopting as standard.
 
\textbf{Offline Usability (D7):} 96.8\% of models nominally support offline use, but this reflects the incidental property that downloadable code works without internet, not deliberate design for field deployment. No model in our corpus supports the pre-cache-and-analyze workflow that would make GeoFM capabilities useful in disconnected field settings by precomputing embeddings for a study region, caching them for local analysis, and synchronizing field-collected labels upon reconnection. As more models adopt cloud-hosted deployment patterns, this dimension will become more discriminative. More importantly, the development of lightweight, edge-optimized downstream tools that operate on cached embeddings represents a significant opportunity for enabling iterative fieldwork scenarios such as adaptive sampling and active learning.

\subsection{Limitations}

We acknowledge several limitations. First, our pilot survey ($N=11$) draws from a single institutional context (conservation and ecology) and cannot claim representativeness across all geospatial application domains. Future work should conduct larger, cross-domain studies. Second, our evaluation captures a snapshot of the landscape in mid-to-late 2025. the GeoFM ecosystem evolves rapidly, and some models may have updated their public releases since our evaluation window. This is inherent to any systematic review and does not invalidate the structural patterns we identify, which reflect field-level norms rather than individual model decisions. Third, the framework evaluates what is publicly documented and available, and as such undocumented capabilities are not captured. We consider this appropriate, as if a capability is not documented, it does not exist from a usability perspective, regardless of what the code may technically support. A corollary is that several dimensions (notably D2 and D3) measure communicated rather than latent capability. A model could support uncertainty estimation in code yet score $T3 = 0$ if this is undocumented. We regard this as appropriate for a usability lens, and it directly motivates our usability-metadata proposal.

Dimension~1 exhibited the lowest inter-rater agreement on the calibration sample (nominal $\kappa = 0.56$, though quadratic-weighted $\kappa = 0.91$ reflecting that disagreements were consistently off-by-one), particularly at the Level~2/3 boundary and this was addressed through discussion before the full coding proceeded. Finally, several dimensions (D5, D6, D7) showed very low variance, limiting their current discriminative utility, though we argue this is itself a finding about the state of the field.
 
\subsection{Future Directions}
 
Three research directions emerge from our analysis. First, the trust gap demands technical innovation: developing uncertainty quantification methods adapted for high-dimensional geospatial embeddings, coupled with visualization interfaces that communicate spatial confidence in ways domain experts can interrogate. Second, our framework should be validated through task-based user studies examining whether the dimensions and levels we define correspond to meaningful differences in actual user performance and satisfaction across diverse geospatial domains (agriculture, urban planning, disaster response). Third, the community should investigate whether the usability metadata standards we propose, specifically structured declarations of framework profiles in model cards, actually influence model selection behavior and development incentives when adopted at scale.
 
\section{Conclusion}
\label{sec:conclusion}
 
We have presented the first usability-centric evaluation framework for geospatial foundation models and applied it to 89 models. Our results reveal systemic accessibility gaps. Over half of models are inaccessible without significant ML expertise, no surveyed model provides uncertainty quantification, and the dominant interaction paradigm (fine-tuning) requires the deepest technical knowledge. These findings, grounded in a pilot survey with domain experts and a systematic evaluation of 89 models, demonstrate that GeoFM developers of the sampled models have optimized for model-centric benchmark performance while underinvesting in the user-centered features that domain experts need to adopt these tools.

The proposed framework's dimensions serve dual functions, as some discriminate between individual models today, helping practitioners identify the most usable systems, while others diagnose systemic gaps that the community must address collectively. Both functions are essential for a field that aspires to translate AI capabilities into real-world environmental impact.
 
Geospatial foundation models represent significant advances in Earth observation. Making them accessible to domain experts, whether ecologists, conservation practitioners, urban planners, or emergency managers, is not just a usability concern. It is, we posit, a prerequisite for these models to fulfill their potential in addressing climate change, biodiversity loss, and environmental degradation.



\section{Author Contributions}
    
\textbf{Robin Young:} Conceptualization, Methodology, Investigation, Formal analysis, Visualization, Writing---original draft, Writing---review \& editing. 

\textbf{Artyom Gabtraupov:} Methodology, Investigation, Data curation, Validation, Writing---review \& editing. 

\textbf{Kenzy Soror:} Methodology, Investigation, Data curation, Validation, Writing---review \& editing. 

\textbf{Srinivasan Keshav:} Supervision, Writing---review \& editing.


\bibliographystyle{elsarticle-num-names} 
\bibliography{cas-refs}

\appendix

\section{Survey Instrument}
\label{app:survey}

The following survey was administered to participants. All participation was voluntary and unpaid, with institutional ethics approval.
 
\subsection*{Section 1: Your Background \& Current Work}
 
\begin{enumerate}
    \item \textbf{What is your primary field(s) of expertise?} \textit{(Select all that apply.)} Ecology / Conservation / Earth Science / Agriculture / Geography / Architecture / Computer Science / Other.
 
    \item \textbf{How frequently do you work with geospatial data (e.g. satellite imagery, GIS data, aerial photos) in your research?} \textit{(Select one.)} Daily / Weekly / Monthly / A few times a year / Rarely or never.
 
    \item \textbf{How many years of experience do you have working with geospatial data?} \textit{(Select one.)} $<1$ / 1--3 / 3--5 / 5--10 / $>10$.
 
    \item \textbf{Please rate your comfort level with the following tools and methods.} \textit{(Scale: 1-No Experience to 5-Expert, one rating per row.)} Desktop GIS Software (e.g. QGIS, ArcGIS) / Cloud Platforms (e.g. Google Earth Engine) / Programming (e.g. Python, R) for data analysis / Command line interfaces and shell scripting.
\end{enumerate}
 
\subsection*{Section 2: Current Workflows \& Pain Points}
 
\begin{enumerate}
\setcounter{enumi}{4}
    \item \textbf{Thinking of a recent geospatial project, how difficult were each of the following steps?} \textit{(Scale: 1-Trivial to 7-A Major Barrier, one rating per row.)} Finding the right dataset for your question / Navigating data portals and their access control systems / Downloading large datasets ($>$100 GB) / Converting data between different formats / Integrating datasets with different projections, resolutions, or timestamps / Setting up the correct software environment / Visualizing intermediate results to check your analysis / Packaging your code and data for reproducibility.
 
    \item \textbf{What is the single biggest bottleneck that slows down your research or causes the most frustration?} \textit{(Open text.)}
 
    \item \textbf{What tools in your current research workflow work very well? What makes your research easier?} \textit{(Open text.)}
\end{enumerate}
 
\subsection*{Section 3: Perspectives on AI Foundation Models}
 
\begin{enumerate}
\setcounter{enumi}{7}
    \item \textbf{How familiar are you with the concept of ``geospatial foundation models'' (e.g. Tessera, Prithvi, Clay, Presto, Aurora, or Google Embedding Fields)?} \textit{(Select one.)} I could train or implement one from scratch / I could use a pretrained model for a new task / I could explain the core concepts but not implement it / I have seen the term and have a general idea / I am not familiar with the term.
 
    \item \textbf{Imagine you could customize a general AI model for a specific task. What would be the biggest barrier?} \textit{(Select one.)} The technical skill required / The effort of preparing my own data / The time to train or fine-tune / The computational cost / I'm not sure.
 
    \item \textbf{To trust the output of an AI model for a scientific publication or policy recommendation, which is most important to you?} \textit{(Select one.)} High accuracy scores on a standard benchmark / The model's reasoning aligns with my domain knowledge / A clear visualization of where the model is confident and uncertain / An explanation of how the model arrived at its conclusion for a specific area.
 
    \item \textbf{If you wanted to train an AI model to recognize a specific phenomenon, what would be your biggest concern?} \textit{(Select one.)} Not knowing what data to include / Not knowing if I have enough training data / Not understanding what architecture to choose / Not knowing what hardware I need / Not being able to tell if training is working / Not knowing when to stop training / Other.
 
    \item \textbf{What foundation models, if any, have you tried? What happened?} \textit{(Open text.)}
\end{enumerate}
 
\subsection*{Section 4: Your Ideal Tools \& Final Thoughts}
 
\begin{enumerate}
\setcounter{enumi}{12}
    \item \textbf{In an ideal world, what kind of tool or AI assistant would make the biggest positive impact on your geospatial research?} \textit{(Open text.)}
 
    \item \textbf{Please rate the potential usefulness of the following AI capabilities for your research.} \textit{(Scale: 1-Not Useful to 7-Extremely Useful, one rating per row.)} Finding similar regions across the globe / Generatively filling in gaps in data / Generating realistic what-if scenarios / Identifying anomalous changes over time.
 
    \item \textbf{When evaluating an AI model, how important are the following properties to you?} \textit{(Scale: 1-Not Important to 7-Critically Important, one rating per row.)} Predictions respect known physical laws / The model understands temporal patterns / Analysis moves fluidly between scales / The model is state-of-the-art on benchmarks / The model is open source.
 
    \item \textbf{If a new, powerful AI model for geospatial analysis was available, in which format would you prefer to access it?} \textit{(Select one.)} Source code on GitHub to adapt and train myself / A pretrained model to deploy and fine-tune / Pre-computed data files (representation vectors) to download / A feature integrated into a platform I already use (e.g. GEE, ArcGIS, QGIS).
 
    \item \textbf{After an AI model provides a result, what methods would you use to validate it?} \textit{(Select all that apply.)} Compare overall accuracy with published benchmarks / Visually spot-check against high-resolution imagery / Compare with my own field data or local knowledge / Check if spatial patterns make ecological/geographical sense / Evaluate on a known challenging area or time period / Other.
 
    \item \textbf{What is your biggest concern about the increasing use of complex, ``black box'' AI models in your field?} \textit{(Open text.)}
\end{enumerate}

\section{Model Evaluation Rubric}
\label{app:rubric}
 
The following rubric was used by both raters to evaluate each model. For each model, raters recorded the model name, version/release year, paper URL, and code repository URL (if available), then assessed each dimension as described below.
 
\subsection*{Dimension 1: Access \& Deployment}
 
\textit{Assign one level (0--5).} If Level 0 is assigned, skip all subsequent dimensions.
 
\begin{itemize}
    \item \textbf{Level 0:} No pretrained weights provided. User must pretrain from scratch.
    \item \textbf{Level 1:} pretrained weights provided, but user must create computational environment from scratch with minimal guidance.
    \item \textbf{Level 2:} Weights plus containerized environment (Dockerfile) or package manifest (requirements.txt).
    \item \textbf{Level 3:} Self-contained notebook or script with sample data for a complete demonstration.
    \item \textbf{Level 4:} Model hosted by provider, accessible through a documented API.
    \item \textbf{Level 5:} Capabilities exposed through a GUI (web app, desktop program, or plugin for existing software).
\end{itemize}
 
\subsection*{Dimension 2: Interaction \& Customization}
 
\textit{Assign one base level AND check all applicable paradigms.}
 
Base Level: (1) Unsupported/Research-Only, or (2) Supported/Usable.
 
Paradigms: (I1) Expert Fine-Tuning / (I2) Prompt-Based Exploration / (I3) On-Demand Embedding / (I4) Interactive retraining / (I5) Field Data Integration.
 
\subsection*{Dimension 3: Trust \& Transparency}
 
\textit{Check all applicable attributes.}
 
(T1) Replicable Benchmarks / (T2) Explainability Methods / (T3) Uncertainty Quantification / (T4) Interactive Validation.
 
\subsection*{Dimension 4: Community \& Support}
 
\textit{Assign one level (1--5).}
 
\begin{itemize}
    \item \textbf{Level 1:} Documentation limited to the academic paper.
    \item \textbf{Level 2:} Detailed README for a technical audience.
    \item \textbf{Level 3:} Dedicated documentation website.
    \item \textbf{Level 4:} Domain-oriented tutorials (blog posts, notebooks, videos).
    \item \textbf{Level 5:} Active community channel (Discord, Slack) plus robust documentation.
\end{itemize}
 
\subsection*{Dimension 5: Scientific Permanence \& Reproducibility}
 
\textit{Assign one base level of openness AND check all applicable policies.}
 
Base Level: (O1) Closed-Source / (O2) Open Core / (O3) Fully Archived with DOI.
 
Policies: (P-S) Service Stability Guarantees / (P-P) Data \& Artifact Portability.
 
\subsection*{Dimension 6: Multilingual Support}
 
\textit{Assign one level (0--2).}
 
Level 0: Monolingual. Level 1: 1--2 additional languages. Level 2: 3+ languages.
 
\subsection*{Dimension 7: Offline Usability}
 
\textit{Assign one level (0--1).}
 
Level 0: Online-only. Level 1: Offline support for meaningful tasks.
 
For each dimension, raters wrote a one-to-two sentence justification.

\section{Per-Dimension Inter-Rater Reliability}
\label{app:irr}
 
Table~\ref{tab:irr_detail} reports agreement statistics for each dimension and attribute on the 20-model calibration sample. We report both Gwet's AC1 and Cohen's $\kappa$. As discussed in Section~\ref{sec:irr}, several attributes exhibit the $\kappa$ paradox. That is, when prevalence is highly skewed (nearly all models receive the same score), $\kappa$'s chance-agreement baseline is inflated, producing low $\kappa$ values despite high observed agreement. Attributes T1 and T2 illustrate this, where raters agreed on 13/14 and 14/14 models respectively, yet $\kappa = 0.00$ for both because nearly all scores were identical. AC1 correctly reflects the high agreement in these cases.
 
\begin{table}[htbp]
\centering
\caption{Per-dimension inter-rater agreement on the 20-model calibration sample. $N$ indicates the number of models rated on that dimension (6 models scored Level~0 on D1 and were not rated on subsequent dimensions). Dimensions marked with $\dagger$ exhibit the $\kappa$ paradox due to extreme prevalence skew.}
\label{tab:irr_detail}
\begin{tabular}{lcccc}
\toprule
\textbf{Dimension / Attribute} & $N$ & \textbf{AC1} & \textbf{$\kappa$} & \textbf{Type} \\
\midrule
D1: Access \& Deployment & 20 & 0.912 & 0.912 & Ordinal \\
D2: Interaction (base level) & 14 & 1.000 & 1.000 & Nominal \\
\quad (I1) Fine-Tuning & 14 & 0.863 & 0.851 & Nominal \\
\quad (I2) Prompt-Based & 14 & 1.000 & 1.000 & Nominal \\
\quad (I3) On-Demand Embedding & 14 & 0.714 & 0.720 & Nominal \\
\quad (I4) Interactive Retraining & 14 & 1.000 & 1.000 & Nominal \\
\quad (I5) Field Data Integration & 14 & 0.886 & 0.811 & Nominal \\
D3: (T1) Replicable Benchmarks$^\dagger$ & 14 & 0.923 & 0.000 & Nominal \\
\quad (T2) Explainability$^\dagger$ & 14 & 0.923 & 0.000 & Nominal \\
\quad (T3) Uncertainty Quantification & 14 & 1.000 & 1.000 & Nominal \\
\quad (T4) Interactive Validation & 14 & 1.000 & 1.000 & Nominal \\
D4: Community \& Support & 14 & 0.887 & 0.888 & Ordinal \\
D5: Permanence (base level) & 14 & 1.000 & 1.000 & Ordinal \\
\quad (P-P) Data Portability & 14 & 1.000 & 1.000 & Nominal \\
\quad (P-S) Service Stability & 14 & 0.912 & 0.632 & Nominal \\
D6: Multilingual Support & 14 & 0.627 & 0.632 & Ordinal \\
D7: Offline Usability & 14 & 1.000 & 1.000 & Nominal \\
\midrule
\textbf{Overall (nominal)} & \textbf{182} & \textbf{0.949} & \textbf{0.922} & \\
\textbf{Overall (ordinal)} & \textbf{62} & \textbf{0.945} & \textbf{0.945} & \\
\bottomrule
\end{tabular}
\end{table}
 
Dimension~1 (Access \& Deployment) exhibited the largest disagreements, though these were consistently off-by-one at the Level~2/3 boundary (quadratic-weighted $\kappa = 0.912$, AC2 $= 0.912$). Nominal $\kappa$ for D1 was lower, reflecting the penalty for any disagreement regardless of magnitude, but weighted statistics correctly capture that these were minor calibration differences rather than fundamental coding disagreements. This was resolved through discussion before the full coding proceeded. 

Dimension~4 (Community \& Support) showed moderate agreement ($\kappa = 0.696$, AC1 $= 0.831$), with two disagreements at the Level~2/3 boundary (detailed README versus dedicated documentation website). All other dimensions and attributes achieved AC1 $\geq 0.86$.

\section{Model Ratings}
\label{app:ratings}
 
Table~\ref{tab:all_models} presents the evaluation profile for all 89 models in the corpus. Models with D1 = 0 (no usable access) were not rated on subsequent dimensions (indicated by --). Paradigms lists the supported interaction paradigms from Dimension~2; Trust lists the supported trust attributes from Dimension~3. D5 reports the base openness level and any applicable policies. 

Here we cite the peer reviewed published versions of each model where possible for canonicity, but corpus collection was based on the most recent available preprints during the evaluation window. As such, ratings reflect publicly available artifacts as of the evaluation window (October 2025 -- January 2026). Where a model's public artifacts changed between the preprint and published versions, our ratings reflect the state observed during the evaluation window, which may differ from the currently linked canonical version.
 
{\scriptsize
\setlength{\LTcapwidth}{\textwidth}
\begin{longtable}{@{}lc l l c l cc@{}}
\caption[Evaluation profiles of all assessed geospatial foundation models]{Evaluation profiles of all 89 geospatial foundation models. D1: Access \& Deployment (0--5). Paradigms: supported interaction modes (I1--I5). Trust: supported trust attributes (T1--T4). D4: Community \& Support (1--5). D5: Scientific Permanence (O1--O3, P-S, P-P). D6: Multilingual (0--2). D7: Offline (0--1).}
\label{tab:all_models} \\
\toprule
\textbf{Model} & \textbf{D1} & \textbf{Paradigms} & \textbf{Trust} & \textbf{D4} & \textbf{D5} & \textbf{D6} & \textbf{D7} \\
\midrule
\endfirsthead

\caption[]{(continued)} \\
\toprule
\textbf{Model} & \textbf{D1} & \textbf{Paradigms} & \textbf{Trust} & \textbf{D4} & \textbf{D5} & \textbf{D6} & \textbf{D7} \\
\midrule
\endhead

\midrule
\multicolumn{8}{r@{}}{\textit{Continued on next page}} \\
\endfoot

\bottomrule
\endlastfoot

A2MAE \citep{zhang2024a2mae} & 0 & -- & -- & -- & -- & -- & -- \\
AIEarth \citep{xu2023aiearth} & 5 & I2, I3, I4, I5 & T1 & 3 & O2, P-S, P-P & 0 & 0 \\
AnySat \citep{astruc2025anysatearthobservationmodel} & 3 & I1 & T1 & 2 & O2 & 0 & 1 \\
BFM \citep{cha2024bfm} & 0 & -- & -- & -- & -- & -- & -- \\
CACo \citep{mall2023caco} & 3 & I1 & T1 & 2 & O2 & 0 & 1 \\
Clay \citep{ClayModel} & 3 & I1, I3, I4 & T1 & 4 & O2, P-S & 0 & 1 \\
CMID \citep{muhtar2023cmid} & 2 & I1, I3 & T1 & 2 & O2 & 0 & 1 \\
Copernicus-FM \citep{wang2025copernicus} & 3 & -- & T1 & 2 & O2 & 0 & 1 \\
CROMA \citep{Fuller2023} & 1 & -- & T1 & 1 & O2 & 0 & 1 \\
Cross-Scale MAE \citep{Tang2023} & 1 & I1 & T1 & 2 & O2 & 0 & 1 \\
CSPT \citep{zhang2022cspt} & 1 & I1, I3 & T1 & 1 & O2 & 0 & 1 \\
CtxMIM \citep{zhang2024ctxmim} & 0 & -- & -- & -- & -- & -- & -- \\
CMC-RSSR \citep{stojnic2021CMCRSSR} & 1 & I1, I3 & T1 & 2 & O2 & 0 & 1 \\
SSLTransformerRS \citep{scheibenreif2022SSLTransformerRS} & 3 & I1, I5 & T1 & 2 & O2 & 0 & 1 \\
DeCUR \citep{wang2024DeCUR} & 2 & -- & T1 & 1 & O2 & 0 & 1 \\
DINO-MC \citep{wanyan2024dinomc} & 1 & I1, I5 & T1 & 2 & O2 & 0 & 1 \\
DINO-MM \citep{wang2022dinomm} & 1 & -- & T1 & 1 & O2 & 0 & 1 \\
DOFA \citep{xiong2024neuralplasticityinspiredmultimodalfoundation} & 3 & I1, I5 & T1 & 2 & O2 & 0 & 1 \\
DynamicVis \citep{chen2025dynamicvisefficientgeneralvisual} & 3 & I1, I3, I5 & T1 & 2 & O2 & 1 & 1 \\
EarthPT \citep{smith2024earthpt} & 2 & -- & T1 & 2 & O2 & 0 & 1 \\
FedSense \citep{tan2025fedsense} & 0 & -- & -- & -- & -- & -- & -- \\
FG-MAE \citep{wang2025fgmae} & 1 & I1 & T1 & 2 & O2 & 0 & 1 \\
FlexiMo \citep{li2025fleximo} & 0 & -- & -- & -- & -- & -- & -- \\
FoMo \citep{bountos2025fomo} & 2 & -- & T1 & 2 & O2 & 0 & 1 \\
Galileo \citep{tseng2025galileo} & 3 & I3 & T1 & 2 & O2 & 0 & 1 \\
GASSL \citep{Ayush2021} & 1 & I3 & T1 & 2 & O2 & 0 & 1 \\
GeCo \citep{Li2022Geco} & 0 & -- & -- & -- & -- & -- & -- \\
GeoKR \citep{Li2022GeoKR} & 0 & -- & -- & -- & -- & -- & -- \\
GeRSP \citep{huang2024gersp} & 2 & -- & T1 & 2 & O2 & 0 & 1 \\
GFM \citep{Mendieta2023} & 2 & I1 & T1 & 2 & O2 & 0 & 1 \\
AlphaEarth \citep{brown2025alphaearthfoundationsembeddingfield} & 5 & I3 & T1 & 4 & O1, P-S, P-P & 0 & 0 \\
Hydro \citep{corley2024hydro} & 3 & -- & T1 & 2 & O2 & 0 & 1 \\
HyperFree \citep{Li2025HyperFree} & 3 & I1, I2, I3, I5 & T1 & 2 & O2 & 0 & 1 \\
HyperSIGMA \citep{wang2025hypersigma} & 3 & I1 & T1 & 2 & O2 & 0 & 1 \\
lal-SimCLR \citep{prexl2023lalSimCLR} & 0 & -- & -- & -- & -- & -- & -- \\
LeMeViT \citep{jiang2024lemevit} & 2 & -- & -- & 2 & O2 & 0 & 1 \\
MA3E \citep{li2024ma3e} & 1 & I1 & T1 & 2 & O2 & 0 & 1 \\
MATTER \citep{Akiva2022} & 1 & -- & -- & 1 & O2 & 0 & 1 \\
MM-VSE \citep{ravirathinam2026knowledge} & 0 & -- & -- & -- & -- & -- & -- \\
MMEarth \citep{Nedungadi2024} & 3 & I5 & T1 & 3 & O2, P-P & 0 & 1 \\
MOSAIKS \citep{rolf2021MOSAIKS} & 4 & I3 & T1, T2 & 4 & O2, P-S & 0 & 1 \\
msGFM \citep{Han2024} & 1 & -- & T1 & 1 & O2 & 0 & 1 \\
MSFE-MMFH \citet{feng2023} & 0 & -- & -- & -- & -- & -- & -- \\
MTP \citep{wang2024mtp} & 3 & I1 & T1 & 2 & O2 & 0 & 1 \\
OFA-Net \citep{xiong2024ofanet} & 0 & -- & -- & -- & -- & -- & -- \\
OmniSat \citep{astruc2024omnisat} & 2 & I1 & T1 & 2 & O2 & 0 & 1 \\
OReole-FM \citep{Dias_2024} & 0 & -- & -- & -- & -- & -- & -- \\
Panopticon \citep{waldmann2025panopticonadvancinganysensorfoundation} & 3 & I1, I3 & T1 & 2 & O2 & 0 & 1 \\
PIEVIT \citep{lu2025pattern} & 0 & -- & -- & -- & -- & -- & -- \\
PIS \citep{an2024pis} & 1 & I1 & T1 & 1 & O2 & 0 & 1 \\
Presto \citep{tseng_lightweight_2024} & 3 & I1, I3, I5 & T1 & 2 & O2, P-P & 0 & 1 \\
Prithvi \citep{schmude2024prithviwxcfoundationmodel} & 5 & I1, I3, I5 & T1 & 5 & O2, P-S, P-P & 0 & 1 \\
Prithvi-EO-2.0 \citep{szwarcman2025prithvieo20} & 3 & I1 & T1 & 2 & O2 & 0 & 1 \\
RingMo \citep{Sun2023} & 2 & I1 & T1 & 2 & O2 & 1 & 1 \\
RingMo-Ae \citep{diao2025ringmoae} & 0 & -- & -- & -- & -- & -- & -- \\
RingMo-Lite \citep{wang2023ringmolite} & 0 & -- & -- & -- & -- & -- & -- \\
RingMo-Sense \citep{yao2023ringmosense} & 0 & -- & -- & -- & -- & -- & -- \\
RingMoE \citep{bi2025ringmoe} & 0 & -- & -- & -- & -- & -- & -- \\
RoMA \citep{wang2025roma} & 2 & I1 & T1 & 2 & O2 & 0 & 1 \\
RS-BYOL \citep{jain2022rsbyol} & 0 & -- & -- & -- & -- & -- & -- \\
RS-DFM \citep{wang2024rsdfm} & 0 & -- & -- & -- & -- & -- & -- \\
RS-vHeat \citep{hu2025rsvheat} & 0 & -- & -- & -- & -- & -- & -- \\
RSP \citep{wang2023rsp} & 1 & I1 & T1 & 2 & O2 & 0 & 1 \\
RVSA \citep{wang2023rvsa} & 2 & I1 & T1 & 2 & O2 & 0 & 1 \\
S2MAE \citep{li2024s2mae} & 0 & -- & -- & -- & -- & -- & -- \\
SAR-JEPA \citep{li2024sarjepa} & 1 & I1 & T1, T4 & 2 & O2 & 1 & 1 \\
SatMamba \citep{duc2026satmamba} & 0 & -- & -- & -- & -- & -- & -- \\
SatDiFuser \citep{jia2025generativegeospatialdiffusionmodels} & 3 & -- & T1 & 2 & O2 & 0 & 1 \\
SatLas \citep{Bastani2023} & 5 & I1, I3 & T1, T4 & 3 & O2, P-S, P-P & 0 & 1 \\
SatMAE \citep{Cong2022} & 2 & I1, I3 & T1 & 2 & O2 & 0 & 1 \\
SatMAE++ \citep{Noman2024} & 1 & I1 & T1 & 2 & O2 & 0 & 1 \\
SatVision-TOA \citep{spradlin2024satvisiontoa} & 3 & I1, I5 & T1 & 2 & O2 & 0 & 1 \\
Scale-MAE \citep{Reed2023} & 2 & I1 & T1 & 1 & O2 & 0 & 1 \\
SeCo \citep{Manas2021} & 3 & I1 & T1 & 2 & O2 & 0 & 1 \\
SeaMo \citep{li2025seamoseasonawaremultimodalfoundation} & 0 & -- & -- & -- & -- & -- & -- \\
SelectiveMAE \citep{wang2025selectivemae} & 2 & I1, I5 & T1 & 2 & O2, P-P & 0 & 1 \\
SenPa-MAE \citep{prexl2024senpamae} & 2 & -- & T1 & 2 & O2 & 0 & 1 \\
SkySense \citep{Guo2024} & 2 & I5 & T1 & 2 & O2 & 0 & 1 \\
SkySense++ \citep{wu2025skysensepp} & 2 & I1, I5 & T1 & 2 & O3 & 0 & 1 \\
SML-FR \citep{dong2024smlfr} & 1 & I1 & T1 & 1 & O2 & 0 & 1 \\
SoftCon \citep{wang2024softcon} & 1 & -- & T1 & 1 & O2 & 0 & 1 \\
SpectralEarth \citep{braham2025SpectralEarth} & 1 & I1 & T1 & 1 & O2 & 0 & 1 \\
SpectralGPT \citep{Hong2024} & 2 & I1 & T1 & 2 & O2 & 0 & 1 \\
SwiMDiff \citep{tian2024swimdiff} & 0 & -- & -- & -- & -- & -- & -- \\
TOV \citep{tao2023tov} & 1 & I1, I5 & T1 & 2 & O2 & 0 & 1 \\
Tessera \citep{feng2025tessera} & 5 & I1, I2, I3, I4, I5 & T1, T4 & 5 & O2, P-S, P-P & 0 & 1 \\
U-Barn \citep{dumeur2024ubarn} & 0 & -- & -- & -- & -- & -- & -- \\
USat \citep{irvin2023usat} & 0 & -- & -- & -- & -- & -- & -- \\
WildSAT \citep{daroya2024wildsatlearningsatelliteimage} & 3 & -- & T1 & 2 & O2 & 0 & 1 \\

\end{longtable}
}

\end{document}